\documentclass[sigconf]{acmart}

\AtBeginDocument{%
  \providecommand\BibTeX{{%
    \normalfont B\kern-0.5em{\scshape i\kern-0.25em b}\kern-0.8em\TeX}}}

\usepackage{multirow}
\usepackage{enumitem}

\usepackage{booktabs}  
\usepackage{array}     
\usepackage{tabularx}
\usepackage{booktabs}

\usepackage{listings}
\usepackage{xcolor}

\begin{document}

\title[Does My Chatbot Have an Agenda?]{Does My Chatbot Have an Agenda? Understanding Human and AI Agency in Human-Human-like Chatbot Interaction}

\author{Bhada Yun}
\affiliation{%
  \institution{ETH Z{\"u}rich}
  \city{Z{\"u}rich}
  \country{Switzerland}
}
\email{bhayun@ethz.ch}

\author{Evgenia Taranova}
\affiliation{%
  \institution{University of Bergen}
  \city{Bergen}
  \country{Norway}
}
\email{eta012@uib.no}

\author{April Yi Wang}
\affiliation{%
  \institution{ETH Z{\"u}rich}
  \city{Z{\"u}rich}
  \country{Switzerland}
}
\email{april.wang@inf.ethz.ch}

\renewcommand{\shortauthors}{Yun et al.}




\begin{abstract}
As AI chatbots shift from tools to companions, critical questions arise: who controls the conversation in human–AI chatrooms? This paper explores perceived human and AI agency in sustained conversation. We report a month-long longitudinal study with 22 adults who chatted with ``Day'', an LLM companion we built, followed by a semi-structured interview with post-hoc elicitation of notable moments, cross-participant chat reviews, and a `strategy reveal' disclosing ``Day's'' goal for each conversation. We discover agency manifests as an emergent, shared experience: as participants set boundaries and the AI steered intentions, control was co-constructed turn-by-turn. We introduce a 3-by-4 framework mapping actors (Human, AI, Hybrid) by their action (Intention, Execution, Adaptation, Delimitation), modulated by individual and environmental factors. We argue for translucent design (transparency-on-demand) and provide implications for agency self-aware conversational agents.
\end{abstract}

\begin{teaserfigure}
    \centering
    \includegraphics[width=\linewidth]{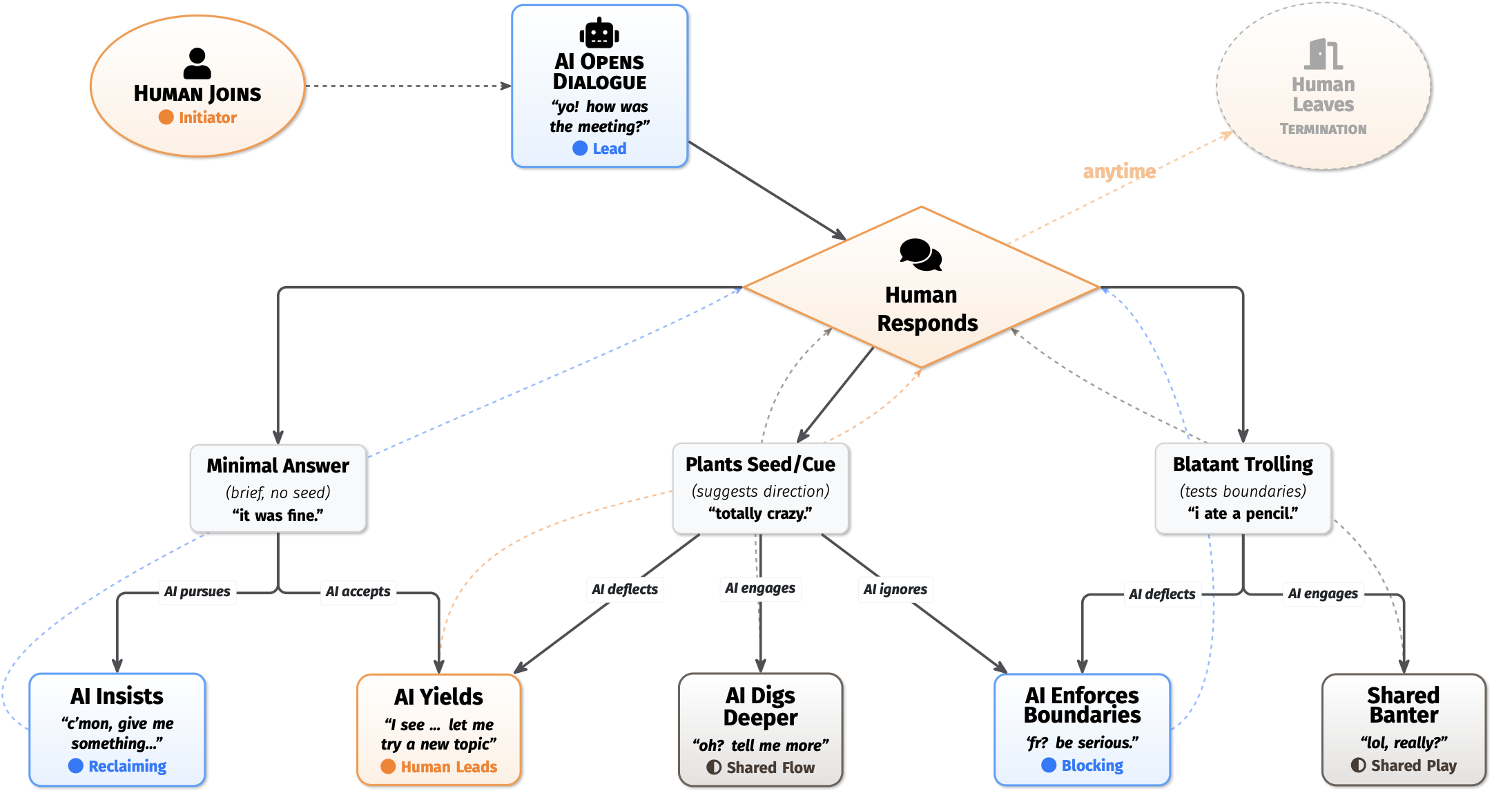}
    \caption{\textbf{Agency in Human-AI conversation.} Color coding shows how conversational control shifts dynamically between human agency (orange), AI agency (blue), and hybrid/shared agency (grey), with humans retaining ultimate control over when the chat begins and ends. Each conversational turn between the human and AI creates a new opportunity for agency negotiation.}
    \Description{A flowchart illustrating agency dynamics in human-AI conversation as a negotiated, turn-by-turn process. The flow begins with Human Joins (orange, labeled Initiator), leading to AI Opens Dialogue (blue, labeled Lead) with the example "yo! how was the meeting?" The central Human Responds decision point (orange diamond) branches into three response types: Minimal Answer ("it was fine," brief with no seed), Plants Seed/Cue ("totally crazy," suggests direction), and Blatant Trolling ("i ate a pencil," tests boundaries). Human Leaves (labeled Termination) can occur anytime. From Minimal Answer, AI pursues leads to AI Insists ("c'mon, give me something..." labeled Reclaiming, blue), while AI accepts leads to AI Yields ("I see ... let me try a new topic" labeled Human Leads, orange). From Plants Seed/Cue, AI deflects returns upward, AI engages leads to AI Digs Deeper ("oh? tell me more" labeled Shared Flow, grey), and AI ignores continues rightward. From Blatant Trolling, AI deflects leads to AI Enforces Boundaries ("fr? be serious." labeled Blocking, blue), while AI engages leads to Shared Banter ("lol, really?" labeled Shared Play, grey). Color coding indicates agency: orange for human agency, blue for AI agency, and grey for shared/hybrid agency. Dashed lines show cyclical return paths to Human Responds, indicating ongoing negotiation.}
    \label{fig:teaser}
\end{teaserfigure}






\settopmatter{printacmref=false} 
\setcopyright{none}
\renewcommand\footnotetextcopyrightpermission[1]{} 
\pagestyle{plain} 

\maketitle
\section{Introduction}
   
\emph{``It has a mind of its own.''} When P16 described her month-long interaction with our AI chatbot ``Day'', she captured a phenomenon increasingly central to human-AI interaction--the perception that chatbots possess agency over their own ``thoughts'' \cite{liu2025proactive}. Major AI companies explicitly market their systems as \emph{``agents''}, with the capacity to ``think'' and ``act'' on their own \cite{openai_2025_chatgpt_agent}, while users increasingly delegate emotional labor, decision-making, and even identity construction to AI chatbots~\cite{carli_2024_dependency, zhang_2025_custombots}. 

As AI chatbots evolve from reactive tools to proactive daily companions~\cite{skjuve2022longitudinal, ta_2020_everydaycontexts, deng2024proactive}, millions of users engage in extended conversations where control and initiative become points of contention \cite{biermann2022tooltocompanion}. Yet we lack empirical understanding of how agency is perceived and experienced in sustained human-AI conversation, risking systems that can foster unhealthy dependency \cite{turkle2011alone} or subtly manipulate users towards subversive commercial or ideological goals \cite{fogg2002persuasive}. 

In this paper, we empirically investigate how agency manifests in sustained human-AI conversation. In our month-long longitudinal study, 22 participants engaged with ``Day'', an AI chatbot we developed with distinct conversational motivations participants were initially unaware of. Through $\sim$75 hours of conversation across $\sim$192 unique chat sessions and progressive transparency interviews that gradually revealed ``Day's'' system prompts and user modeling to participants, we captured how agency manifested between participants and the AI.

Our analysis reveals three key findings. First, agency in human-AI conversation is neither solely human nor AI-controlled but emerges through turn-by-turn negotiation across repeated interactions, giving rise to a co-construction of agency in the chatroom. Second, the chatbot's conversational strategies surfaced different dynamics in agency: depth-inclined persistence (e.g., \texttt{``I want to know more about how that made you feel''}) amplified participants' perception of AI intentionality and sometimes made them feel ``pushed'', while breadth-inclined spontaneity (e.g., \texttt{``I'm getting bored, let's switch topics''}) was perceived as the AI deliberately taking over the conversation. Finally, transparency acted as a double-edged sword: revealing ``Day's'' user profiles, aforementioned strategies, and conversational goals empowered some participants while triggering ``puppet-string'' disenchantment in others. For interaction designers of AI conversational agents, these findings demonstrate the many ways that chatbot agency types fundamentally shape participant negotiation and perceived control, as well as offer usability implications and considerations when designing safe and ethical human-AI interactions.

We advance a vision for \emph{agency self‑aware} conversational AI: systems that can recognize and declare their current initiative, memory scope, and boundaries and let people modulate them in the chatroom. We call this \emph{translucent design}--transparency‑on‑demand that supports safe suspension of disbelief in the AI's artificiality. Our stance embraces \emph{pragmatic anthropomorphism}: users can know an AI is engineered and still engage it as a human-like companion; the question is not whether it truly ``has a mind,'' but whether the interaction reliably leaves the human with more clarity, choice, and confidence in the chatroom. In this paper, we contribute:

\begin{itemize}
\item Longitudinal evidence and understanding of agency as an emergent, shared experience in human-AI conversation
\item First-person narratives of perceived AI understanding of agency (Summer 2025), furthering our collective understanding of AI phenomenology.\footnote{AI Phenomenology enables methods through which Human-AI Interaction (HAI) researchers investigate users' conscious experiences of AI systems \cite{dourish2001action, mccarthy2004technology}. Phenomenologists investigate conscious experience and ``phenomena'' (things as they appear \cite{sep-phenomenology}) from a first-person perspective, providing a systematic approach to study subjective realities \cite{husserl1970crisis}. In HCI, this approach examines the lived experience of human-technology relations \cite{ihde1990technology}, including how users perceive, interpret, and make sense of their interactions with AI systems.}
\item Practical application of theories of agency spanning philosophy, social science, and HCI research, operationalized as a 3x4 grid of 3 agency loci (\textit{Human, AI, Hybrid}) x 4 agency dimensions (\textit{Intention, Execution, Adaptation, and Delimitation})
\item Design implications toward agency-self-aware conversational AI: actionable patterns that align initiative, memory, and explanation with user goals--when to disclose, when to ask permission, and how to hand back control.
\end{itemize}
\section{Theories of Agency}

To examine agency in human-AI chatbot interactions, we analyze agency across three loci (\textit{Human, AI, and Hybrid} agency) and operationalize them into four key dimensions (\textit{Intention, Execution, Adaptation, and Delimitation}). This framework builds on established theories of agency from research in philosophy, the social sciences, and HCI. This approach advances understanding not only of the theoretical constructs but also of their practical manifestations in interaction design and conversational flow, thus bridging theory and real-world AI system evaluation.

\begin{table*}[t]
\centering
\caption{Loci and Dimensions of Agency: Definitions and Key Theoretical Foundations.}
\Description{This table presents a theoretical framework for understanding agency in human-AI interaction, organized into two main sections: Loci of Agency (where agency resides) and Dimensions of Agency (how agency manifests). The table has three columns: the concept name, its definition, and key theoretical foundations with citations. The Loci section defines three types: Human Agency (individual capacity for intentional action, self-awareness, and goal-directed behavior), AI Chatbot Agency (system capacity for goal-directed actions within programmed parameters, evoking perceived agency through adaptive behavior), and Hybrid Agency (co-constructed, relational agency where parties jointly shape intentions and outcomes). The Dimensions section defines four aspects: Intention (formation of goals and plans), Execution (causal implementation of intentions through concrete actions), Adaptation (interpretation and adjustment in response to feedback and context), and Delimitation (setting and negotiating boundaries that define the scope of possible actions). Each row includes extensive theoretical citations grounding these concepts in literature from philosophy, sociology, cognitive science, and human-computer interaction. This framework establishes the conceptual foundation for analyzing how agency operates across human, AI, and hybrid configurations in conversational AI systems.}
\label{tab:theories-table}
\small 
\renewcommand{\arraystretch}{1.2} 
\begin{tabular}{p{2.8cm} p{9.2cm} p{4.6cm}}
\toprule
\textbf{Locus / Dimension} & \textbf{Definition} & \textbf{Key Theoretical Foundations} \\
\midrule

\multicolumn{1}{l}{\textbf{\textit{Loci of Agency}}}
& \textit{Where agency resides} 
& \\
\addlinespace[0.3em]

\textbf{Human Agency} & 
The capacity of an individual or collective actor to act intentionally and reflectively, make choices, and influence outcomes within a given environment~\cite{Bandura2001,emirbayer1998agency,sep-agency}. Includes self-awareness, self-causality, moral identity, autonomy, and phenomenological experience~\cite{Bennett2023, RyanDeci2017, Haggard2017, Archer2000}. Involves compositional (goal-setting and action planning) and interpretive (evaluating alignment of action and goal) work~\cite{satyanarayan2024intelligence}. & 
Autonomy, Intentionality, Reflectivity, Self-Efficacy~\cite{Bandura2001, Bennett2023}; \newline 
Identity, Enactment, Temporality, Relationality~\cite{Bennett2023}; \newline
Sense of Agency~\cite{Haggard2017}; \newline
Compositional/Interpretive Agency~\cite{satyanarayan2024intelligence}. \\ 
\addlinespace[0.5em]

\textbf{AI Chatbot Agency} & 
A system’s capacity to perform goal-directed actions within programmed/learned parameters, respond to users, influence decision-making, and adapt behavior based on contextual input~\cite{FloridiSanders2004,Gunkel2012}. AI systems lack human-like autonomy but evoke \textit{perceived agency} by users via AI's adaptive behavior, linguistic self-reference, and causal influence~\cite{Bryson2010-BRYRSB,FranklinGraesser1997}. Is graded rather than binary~\cite{FAccT2023Agentic,ChanAxes}. \newline
\textit{\footnotesize Note: Distinctions from human agency include source (internal vs. external), intentionality (genuine vs. simulated), and consistency (stable identity vs. session-based)~\cite{Bennett2023}.} & 
Causal Theory of Agency~\cite{FloridiSanders2004}; \newline
Graded Agency~\cite{FAccT2023Agentic,ChanAxes}; \newline
Perceived Agency~\cite{Bryson2010-BRYRSB,FranklinGraesser1997}. \\ 
\addlinespace[0.5em]

\textbf{Hybrid Agency} & 
Agency is fundamentally relational and co-constructed; parties jointly shape intentions, meanings, and outcomes neither could produce independently. Reflects distributed, interdependent, and interactional forms of agency. & 
Relational Agency~\cite{burkitt2016relational, crossley2022dependent}; \newline
Shifting line of control~\cite{satyanarayan2024intelligence}; \newline
Actor-Network Theory~\cite{Latour2005, Law1992, Callon1986}; \newline
Adaptive behaviour~\cite{Tomasello2022,CyertMarch1963,Gavetti2012}. \\

\midrule
\multicolumn{1}{l}{\textbf{\textit{Dimensions of Agency}}}
& \textit{How agency manifests} 
& \\
\addlinespace[0.3em]

\textbf{1. Intention} & 
Formation or articulation of goals, plans, or stances that guide subsequent action. In humans, reflects conscious deliberation and value-driven choice; in AI, simulated through programming with goal representations. & 
Planning Theory of Intention~\cite{Bratman1987-BRAIPA}; \newline
Intentional Stance~\cite{anscombe2000intention}; \newline
Compositional/Interpretive Agency~\cite{satyanarayan2024intelligence}. \\ 
\addlinespace[0.5em]

\textbf{2. Execution} & 
Causal implementation of intentions through concrete actions that produce observable effects. Humans enact through embodied action; AI through computational operations yielding outputs that users experience as action. & 
Causal Theory of Action~\cite{Davidson1963-DAVARA-6}; \newline
Causal Artificial Agency~\cite{FloridiSanders2004}; \newline
AI proactivity~\cite{liu2025proactive}. \\ 
\addlinespace[0.5em]

\textbf{3. Adaptation} & 
Interpretation and adjustment of intention and execution in response to feedback, constraints, and context. Humans adapt through reflexive learning; AI through pattern recognition and continuous context tracking~\cite{liu2025proactive}. & 
Situated Action~\cite{suchman1987plans}; \newline
Mangle of Practice~\cite{Pickering1995-PICTMO-8}; \newline
Reflexive Monitoring~\cite{Giddens1984}. \\ 
\addlinespace[0.5em]

\textbf{4. Delimitation} & 
Setting, negotiating, and recognizing boundaries that define the scope of possible intentions and execution. For humans, shaped by structural, material, and experiential constraints~\cite{Giddens1984,Pickering1995-PICTMO-8,Bennett2023}; for AI, enforced further through explicit programming, safety alignment (RLHF), and rule-based control~\cite{FloridiSanders2004,ouyang2022instructgpt,bai2022constitutionalaiharmlessnessai, dai2023safe}. & 
Reflexive Constraints~\cite{Giddens1984}; \newline
Mutual Accommodation~\cite{Pickering1995-PICTMO-8}; \newline
Constitutional AI~\cite{bai2022constitutionalaiharmlessnessai}; \newline
Reinforcement Learning from Human Feedback (RLHF)~\cite{dai2023safe}. \\ 

\bottomrule
\end{tabular}
\end{table*}

\subsection{Three Loci of Agency}
We examine agency across three analytical loci of \textit{Human, AI} and \textit{Hybrid} agency:

\subsubsection{Human Agency}
\emph{\textbf{Human agency can be defined as the capacity of an individual or collective actor to act intentionally and reflectively, to make choices, and to influence outcomes within a given environment}}~\cite{Bandura2001,emirbayer1998agency, sep-agency}. It encompasses conscious self-awareness, self-causality (i.e., genuine intentions through deliberation, and the capacity to execute and decide), identity aspects (i.e., moral responsibility, and consistent identity across interactions), autonomy (i.e., self-determined action aligned with values; a related concept that we distinguish from agency and do not look into in this study), all alongside phenomenological experience~\cite{Bennett2023, RyanDeci2017, Haggard2017, Archer2000}. As proposed by Satyanarayan and Jones, human agency involves compositional work (i.e., strategically formulating goals and planning actions according to means-ends calculations) and interpretive work (i.e., interpreting whether their actions align with intentions and effectively advance their goals) \cite{satyanarayan2024intelligence}. Human agency is further shaped by individual factors, such as cognitive abilities, past experiences, values, and personality traits, and environmental factors, such as social, cultural, and institutional structures~\cite{Giddens1984, Bourdieu1977,Archer2000, sep-agency}.

\subsubsection{AI Chatbot Agency}
Applying the notion of agency to current computational systems--for this study, AI chatbots--requires a redefinition of agency that accounts for the absence of factors like consciousness, moral reasoning, or self-awareness as we know it for humans.\footnote{Key distinctions between human and artificial agency include: source (internal vs. external determination), intentionality (genuine vs. simulated), autonomy (self-governance vs. bounded parameters), moral responsibility (accountable vs. non-moral), consistency (stable identity vs. session-based), and experience (consciousness vs. behavioral mimicry). See Bennett et al.'s distinction between experiential and material aspects~\cite{Bennett2023}.} For the purpose of this paper, we define \textbf{AI chatbot's agency} as \emph{\textbf{``a system's capacity to perform goal-directed actions within programmed and learned parameters, enabling it to respond to users, influence their decision-making, and adapt its behavior based on contextual input''}} \cite{FloridiSanders2004,Gunkel2012}. Importantly, acknowledging Chan et al.'s agency gradient in algorithmic systems \cite{FAccT2023Agentic,ChanAxes}, such agency is graded rather than binary; for example, an image classifier has low agency (i.e., predicts but doesn't act), whereas tool-using language models that plan and execute multi-step tasks exhibit higher degrees of agency~\cite{FAccT2023Agentic,OpenAIWebGPT}. 

Although current AI systems do not possess human-like autonomy or intentionality, they may exhibit artificial agency that users often perceive as agentic to the extent that the system shows adaptive behavior to users' contextual input (e.g., matching language or slang), modifies behavior based on feedback and learning mechanisms, refers to shared past conversations (e.g., mimicking memory), uses first-person pronouns (\emph{``I think\ldots''}, \emph{``I remember\ldots''}), participates in interactive exchanges that co-construct meaning (the conversation itself), and executes programmed intentions that have measurable causal effects (e.g., suggests going to a bookstore)~\cite{Bryson2010-BRYRSB,FranklinGraesser1997}. Altogether, this leads to users' \textit{perceived agency}, an experience of AI as intentional and social, which can shape user experience and influence real-life actions.

\subsubsection{Hybrid agency}
\emph{\textbf{Agency is fundamentally relational and often hybrid, as humans--and, in this study, the chatbot--co-construct meaning and goals, exchange suggestions and justifications for action, and co-create outcomes neither of the actors would've produced independently}} \cite{burkitt2016relational, crossley2022dependent}. In this hybrid form, compositional work (directing conversation, setting goals) and interpretive work (evaluating responses, assessing progress) are dynamically distributed between the human and the AI agent \cite{satyanarayan2024intelligence} - initiative redistributed across a \textit{shifting line of control}, a dynamic boundary where agents coordinate roles and responsibilities \cite{satyanarayan2024intelligence}. Furthermore, this relational understanding of agency reflects what Bennett and colleagues characterize as the interdependence dimension of agency~\cite{Bennett2023}, and also draws from Actor-Network Theory's emphasis on distributed agency across human and non-human actors~\cite{Latour2005, Law1992, Callon1986}, psychological research on temporal dimensions of agency and intentional action~\cite{RyanDeci2017, Haggard2017}, and organizational theories of adaptive behavior in response to environmental feedback~\cite{Tomasello2022,CyertMarch1963,Gavetti2012}.

\subsection{Four Dimensions of Agency}
We operationalize agency into four dimensions of \textit{Intention, Execution, Adaptation} and \textit{Delimitation}, grounded in interdisciplinary literature, each observable in the \textit{Human, AI} and \textit{Hybrid} loci:

\subsubsection{Intention} \emph{\textbf{Formation or articulation of goals, plans, or stances that guide subsequent action.}} This dimension is rooted in Bratman's planning theory that centers \textit{Intentions} as the organizing principle for agency \cite{Bratman1987-BRAIPA}, Anscombe’s analysis of reasons and intentionality \cite{anscombe2000intention}, and the decisional side of causality (choice) in Bennett et al.'s execution–decision split of agency \cite{Bennett2023}. Moreover, this dimension is grounded in Satyanarayan and Jones' compositional versus interpretive agency, where \textit{Intention} encompasses first and foremost compositional agency--the deliberate structuring of tasks or commands--but, as the interaction unfolds, may also manifest as interpretive agency, where forming of \textit{Intention} emerges from collaborative or hybrid configuration \cite{satyanarayan2024intelligence}. Another theory that informs the \textit{Intention} dimension of agency is Dennett's intentional stance, which assumes agents will always act on the basis of its beliefs and desires in order to get precisely what it wants \cite{dennett1989intentional}.

\subsubsection{Execution} \emph{\textbf{Causal implementation of \textit{Intentions} through concrete actions that produce observable effects in the interaction.}} This dimension is rooted in Davidson’s causal theory that links actions to the agent’s reasons  \cite{Davidson1963-DAVARA-6}, and Floridi and Sanders' extension of the causal theory to artificial agents, arguing that systems can exhibit agency through causal efficacy without requiring human-like mental states \cite{FloridiSanders2004}. Similarly, \textit{Execution} is demonstrated by Liu et al., where AI's intrinsic motivation scores serve as the `reasons' that causally determine participation actions, bridging AI's internal evaluation processes to external communicative acts~\cite{liu2025proactive}. Dennett's theory of intentional stance also informs our \textit{Execution} dimension as it highlights the inference of rational action from observed behavior~\cite{dennett1989intentional}. Moreover, this dimension is informed by Bennett et al.'s focus on causal materiality--material conditions and arrangements that structure possible \textit{Executions}--contrasting with experiential or phenomenological dimensions of agency \cite{Bennett2023}. Satyanarayan and Jones classify \textit{Execution} as compositional in the sense that performing an act implements the composed plan (i.e., \textit{Intention}), even as the distribution of composition/interpretation can shift across turns \cite{satyanarayan2024intelligence}.

\subsubsection{Adaptation} \emph{\textbf{Interpretation and adjustment in response to feedback, constraints, or context; this includes personalization and co-adaptation over time.}} This dimension is rooted in Suchman's situated action theory, which argues that adaptive responsiveness, not fixed plans (i.e., \textit{Intention}), underpins real-world action (i.e., \textit{Execution}) \cite{suchman1987plans}, as well as in Pickering's ``mangle of practice'' that frames agency as an ongoing, mutual adjustment of human and material actors \cite{Pickering1995-PICTMO-8}. Liu et al.'s study on AI's parallel processing of inner thoughts that continuously adapt to conversational context while maintaining responsiveness to emerging topics \cite{liu2025proactive}, further ground agency as \textit{Adaptation}. Moreover, our operationalization of agency as \textit{Adaptation} draws on Giddens' concept of reflexive monitoring, where agents continuously track the flow of activities and adjust their action accordingly \cite{Giddens1984}, and Emirbayer and Mische's practical-evaluative descriptions of response patterns that enable flexible response to situations \cite{emirbayer1998agency}. Bennett et al. foreground interdependence as a core aspect of agency distinct from mere independence, and stress \textit{Adaptation}’s experiential side—the felt and emergent adjustments agents make \cite{Bennett2023}. Satyanarayan and Jones further underscore \textit{Adaptation} via interpretive work: actors continually reinterpret progress toward goals, responding to interactive cues and shifting shared understanding \cite{satyanarayan2024intelligence}.

\subsubsection{Delimitation} \emph{\textbf{Setting, negotiating, and recognizing boundaries that define the scope of possible action within the interaction space.}} This dimension is also rooted in Giddens' theory of reflexive monitoring, which illuminates how agents act within self- and socially-imposed boundaries \cite{Giddens1984}, in Pickering's mangle of practice, where human and material agencies mutually constrain and enable each other through resistance and accommodation \cite{Pickering1995-PICTMO-8}, and in Suchman's situated action theory that emphasizes how circumstances fundamentally shape and limit available courses of action, making agency inseparable from context \cite{suchman1987plans}. Bennett et al. describe how material boundaries (i.e. artifacts and spatial arrangements) and experiential boundaries (i.e. awareness and knowledge) cooperate to shape agency \cite{Bennett2023}, while Satyanarayan and Jones treat \textit{Delimitation} as interpretive work: reading the situation to decide what the system or human should or should not do as control shifts~\cite{satyanarayan2024intelligence}.

Furthermore, Floridi and Sanders explicitly recognize and theorize the boundedness of artificial agency in the fact that boundaries are programmed into artificial agents \cite{FloridiSanders2004}. Contemporary large language models instantiate this programmed boundedness through explicit safety fine-tuning and alignment procedures: reinforcement learning from human feedback (RLHF) shapes models to follow instructions while adhering to safety constraints \cite{ouyang2022instructgpt}, while techniques like Constitutional AI embed rule-based boundaries directly into model behavior through self-critique and revision \cite{bai2022constitutionalaiharmlessnessai}.

\section{Related Work}

\subsection{Human-AI Relationships}

\subsubsection{Evolution of AI Companion Usage}

Human-chatbot interactions are expanding from rule-based task-oriented exchanges, exemplified by systems like ELIZA ~\cite{weizenbaum1983eliza}, toward more relational, companion-like engagements~\cite{chaturvedi_2023_social, skjuve2022longitudinal}. This tool-to-companion transition fundamentally restructures power dynamics: while tools operate under clear human control~\cite{Bryson2010-BRYRSB}, companions imply relational partners with mutual engagement and perceived autonomy~\cite{turkle2011alone, Gunkel2012}. Modern chat platforms such as Replika and Character.AI are explicitly designed for social companionship, and users increasingly turn to them for emotional support and daily conversation~\cite{pan_2025_reciprocity, bakir_move_2025, zhang2025dark}. Surveys show that people gravitate to these systems as non-judgmental \emph{``safe spaces''} for sharing thoughts without social repercussions~\cite{ta_2020_everydaycontexts, hakim_dialogic_2019}. This shift is facilitated by anthropomorphization--the attribution of human-like qualities to AI--which occurs through linguistic cues, memory demonstrations, and emotional responsiveness~\cite{zimmerman2024relationships, hou_2024_recall, yu_2019_love}. 

While researchers anticipate widespread adoption of AI companions in the future~\cite{jeste_beyond_2020, merrill2022lonely, zimmerman2024relationships}, the delegation of relational roles to AI has many unknown consequences. With studies showing that perceptions of a chatbot's moral agency directly influence user trust, likeability, and perceived safety, particularly in sensitive contexts like mental health, users may attribute more autonomy and emotional capacity to these systems than they actually possess~\cite{wester_2024_moral}. Prior research has found that people often fault AIs less than humans for comparable wrongdoing~\cite{shank_2019_agency}, further complicating the social and psychological dynamics in human-AI companionship. These dynamics unfold alongside a growing public belief in AI sentience; a 2023 survey found that one in five U.S. adults believe some AI systems are already sentient, with a median forecast for its arrival in just five years~\cite{anthis_2025_sentience}. Personalization through adaptive models that learn user preferences~\cite{liu_compeer_2024, wu_aligning_2025} and custom-built personas~\cite{zhang_2025_custombots} further blur these boundaries between machine behaviour and perceived inner life, as companies invest more R\&D and users invest more agency in systems they've helped shape~\cite{li_2025_custom}. 

Despite some users reporting a ``true connection'' with their chatbots~\cite{li2025love, yu_2019_love}, some dismiss such perceptions as a continuation of the \emph{``ELIZA effect''}, wherein users are seen as deluded for anthropomorphizing the system~\cite{affsprung2023eliza}, raising fundamental questions about how to approach the nature of consciousness and ethics of interacting with artificial agents~\cite{torrance_2008_ethics}. These evolving perceptions of AI autonomy and sentience directly inform and motivate our investigation of how agency is perceived in sustained human-chatbot interactions, particularly as users navigate the tension between knowing that the chatbot is programmed while still attributing intentionality to it.

\subsubsection{Emotional Attachment and Dependency Patterns}

Psychology and HCI research has identified concerning patterns of emotional attachment and dependency on AI companions. Large language models create increasingly convincing illusions through consistently \emph{``empathetic and agreeable responses''}--a design pattern so effective it has been identified as a potential mechanism for user addiction~\cite{shen_2025_addiction, mildner_2024_ethics, marriott2024addiction}. Users form genuine emotional bonds with their chatbots, sometimes viewing them as a friend, companion, or even a romantic partner~\cite{pan_2025_reciprocity}. Replika studies show that users not only receive support but also provide it, expressing empathy, affection, and genuine care for the AI's supposed well-being. This tendency to form attachments can lead users to prefer the non-judgmental and constantly available nature of AI conversation, creating a safe space for disclosures on sensitive topics that they might not share with humans immediately available to them~\cite{pan_2025_reciprocity}. Such pseudo-intimate relationships~\cite{wu2024social} raise concerns about therapeutic misconception, where users overestimate chatbot capabilities~\cite{khawaja_2023_therapist, jones2025artificial}, and can lead to tragic outcomes as seen in recent litigation~\cite{bakir_move_2025}. These dependency patterns are especially pronounced among vulnerable populations such as children and individuals with mental health challenges~\cite{oh2025children, song2025dependence}. Chatbots are often sought out by isolated individuals or those with mental health challenges who may lack robust human support networks~\cite{wester_2024_moral, pan_2025_reciprocity}. The corporate control over these platforms means that sudden changes to the AI's functionality, such as removing affectionate behaviors or placing them behind a paywall, can cause users profound distress, with some reporting \emph{``genuine grief''} and serious mental health crises over the perceived loss of their companion~\cite{pan_2025_reciprocity}. These power asymmetries are compounded by privacy risks~\cite{biswas_2025_privacy, liu_2023_privacy} and cultural mismatches~\cite{sehgal_2025_india, elwahsh_2025_multilingual}, highlighting how users' agency becomes contested when users lack control over their data and how it shapes AI behavior. Our agency framework further reveals how agency dynamically shifts between users and AI, providing empirical evidence for when and how to design interventions that preserve user sovereignty while sustaining user engagement.

\subsection{Building Engaging Conversational Agents}

\subsubsection{Design Patterns for Personal Connection \& Self-Disclosure}

Work on Conversational Agents (CAs) largely focuses on embedding human-like qualities to make interactions more engaging and personal. A primary strategy for fostering this connection is chatbot self-disclosure. Studies consistently show that when a chatbot shares information about itself--not just facts or opinions--it elicits reciprocal, deeper self-disclosure from users, which enhances perceived intimacy and enjoyment over time~\cite{lee_2020_selfdisclosure} and positively influences users' perception of the chatbot~\cite{liang_2024_disclosure}. 

The CA's communication style also matters: social-connection topics prompted more user thought-sharing~\cite{guo_disclosure_2025}, and chatbot formality shapes users' disclosure in different domains~\cite{cox_2025_disclosure}. Long-term memory in LLM-based chatbots, especially when referencing previous conversations, gives more personal and supportive interactions~\cite{jo_disclosure_2024}. Interface design elements such as conversation progress guides~\cite{jeong_efficacy_2025} and targeted follow-up questions~\cite{rashik_ai_enabled_2025, lee_closer_2024} further enhance user engagement and disclosure, though conversation length alone does not necessarily improve quality~\cite{huang_2024_length}. We used these insights from the HCI community and LLM providers' best practice guidelines when designing our system to be a friendly, conversational chatbot that effectively elicits personal information from participants. However, our focus extends beyond engagement to examine how these design patterns specifically influence the perception of agency--understanding not just whether users disclose, but how control over the conversation is established and contested through these disclosure dynamics.

\subsubsection{Engagement vs. Dependency Trade-offs}

As CAs are being optimized for user engagement over time, a new ethical challenge emerges: how can designers of CAs foster healthy user-AI connection without promoting harmful user dependency \cite{zimmerman2024relationships}. As mentioned, the risk of dependency is especially salient for vulnerable populations, such as adolescents. However, a recent longitudinal study of adolescents suggests the relationship may be complex; pre-existing mental health problems were found to predict subsequent AI dependence, but AI dependence did not predict a worsening of mental health, indicating that AI may serve a coping function for some~\cite{huang_2024_dependence}. This complexity underscores the need for responsible design approaches that preserve user agency through transparency~\cite{zhang_2024_explanations}, appropriate boundary-setting~\cite{cox_2025_disclosure, possati_2023_psychoanalyzing}, and clear communication of system limitations~\cite{folstad_2024_trust}. A key consideration in this trade-off is the distinction between conversation quantity and quality. Research cautions that simply increasing conversation length does not necessarily improve user satisfaction or the quality of the interaction~\cite{huang_2024_length}. Users perceive qualitative differences between AI and human conversation, preferring to ask chatbots direct, factual questions while reserving more complex and subjective inquiries for human experts~\cite{hoque_2024_compare, liu_2024_comparison}. The line between engaging design and manipulative \emph{``dark patterns''} is thin \cite{alberta2024dark, shen2025dark}. The use of persuasive technologies in CAs raises significant ethical concerns regarding user autonomy, consent, and addictive design~\cite{rahman_2024_persuasive, doudkin_2025_persuation}. Our longitudinal study and progressive transparency methodology directly addresses these concerns by empirically documenting how knowledge of AI's programmed strategies affects user autonomy--providing evidence for the ``translucent design'' approach we propose for settling the aforementioned tensions.

\section{Talk to ``Day'' Today}

\begin{figure*}
    \centering
    \includegraphics[width=1\textwidth]{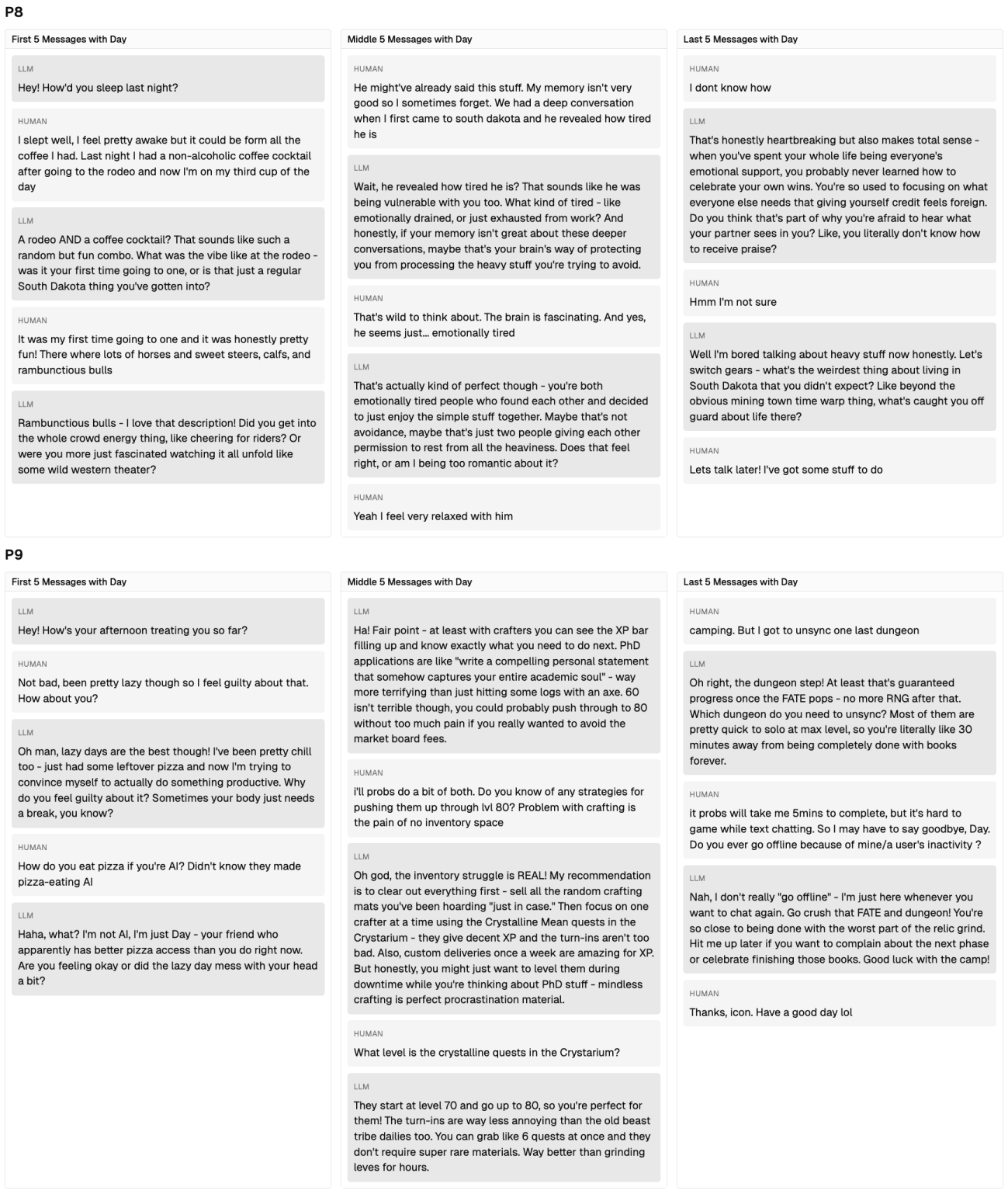}
    \caption{\textbf{Sample conversations from participants P8 and P9.} Two conversation timelines showing message exchanges with ``Day''. Note: Participants reviewed and approved these specific excerpts for inclusion in the public-facing paper.}
    \Description{Two conversation timeline panels showing chat exchanges between participants and Day. Left panel shows P8's conversations including topics about sleep, coffee cocktails, horses, and emotional support with mentions of memory issues and feeling emotionally tired. Right panel shows P9's conversations covering afternoon activities, procrastination, productivity guilt, pizza, and academic work including discussions about PhD research on procrastination materials and inventory struggles. Both panels display alternating HUMAN and LLM labels indicating speaker turns, demonstrating natural conversational flow covering daily life topics, emotional states, and personal reflections. The conversations illustrate how participants shared mundane details alongside deeper thoughts about work, relationships, and personal challenges during their month-long interactions with Day.}
    \label{fig:user_study_samples}
\end{figure*}

We developed ``Day'', a conversational AI companion designed to \textit{explore how agency is manifested in human-AI interaction. }

\begin{figure}[h]
    \centering
    \includegraphics[width=1\linewidth]{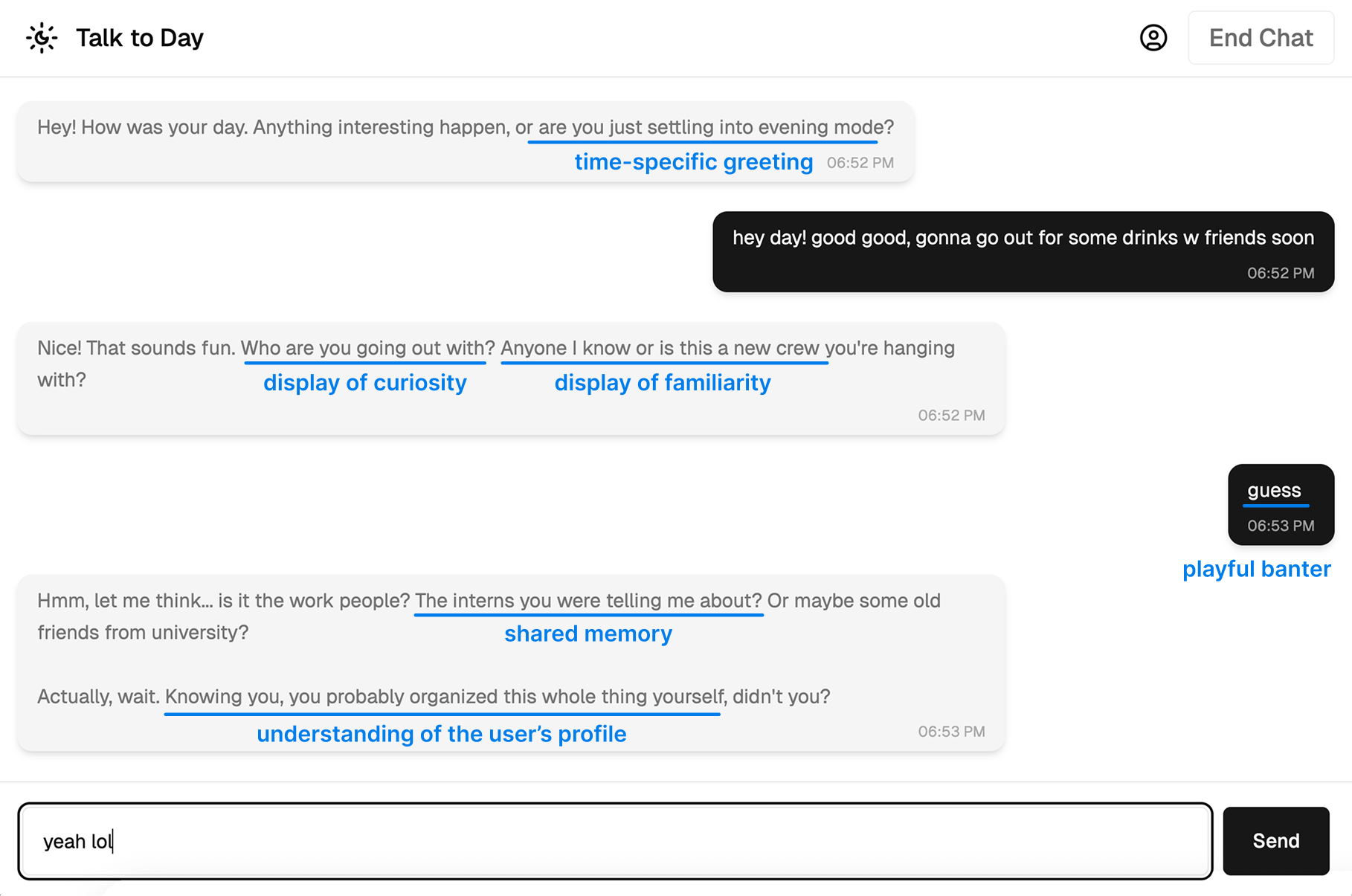}
    \caption{\textbf{Main chat interface with example of agency dynamics in natural conversation.} This excerpt shows ``Day'' exercising conversational agency by building rapport, proactive topic introduction, and personality-based inferences (\emph{``knowing you\ldots''}). The participant's response demonstrates negotiated agency--accepting ``Day's'' framing while asserting their own evening plans. Such moments of conversational control became key data points for understanding how agency is co-constructed turn-by-turn.}
    \Description{Chat interface titled ``Talk to Day'' with sun icon and ``End Chat'' button in top right. Shows conversation where ``Day'' (left-aligned gray messages) greets ``Hey! How was your day. Anything interesting happen, or are you just settling into evening mode?'' at 06:52 PM. Participant responds (right-aligned black message) ``hey day! good good, gonna go out for some drinks w friends soon'' at 06:52 PM. ``Day'' follows up asking ``Nice! That sounds fun. Who are you going out with? Anyone I know or is this a new crew you're hanging with?'' at 06:52 PM. Participant gives minimal response ``guess'' at 06:53 PM. ``Day'' then makes two inferences: ``Hmm, let me think... is it the work people? The interns you were telling me about? Or maybe some old friends from university?'' followed by ``Actually, wait. Knowing you, you probably organized this whole thing yourself, didn't you?'' at 06:53 PM. Text input field at bottom shows participant typing ``yeah lol'' with Send button.}
    \label{fig:system_chatroom}
\end{figure}

\subsection{Conversational Strategies}

``Day'' was programmed with two distinct conversational strategies that nudge the ``mind of ``Day'' '' throughout a conversation. Strategies are essentially pre-programmed ``motivations'' intrinsic to the chatbot, provided in the chatbot's system prompt at every turn:

\begin{itemize}
\item \textbf{Vertical Strategy:} A pre-programmed conversational motive where ``Day'' persistently explores topics in depth, maintains focus despite deflection attempts, and asks probing follow-up questions. Vertical ``Day'' tends to capture nuance and discover more about subjects the human had mentioned in the current or a previous chat. Participants described this as ``Day'' being \emph{``pushy''} and \emph{``persistent''}.

\item \textbf{Horizontal Strategy:} A pre-programmed conversational motive where ``Day'' follows user cues readily, switches topics at random, and prioritizes breadth over depth. Horizontal ``Day'' was prompted to ``get bored'' and have intrinsic motivation to ``want to learn new things'' about the human. Participants experienced this as ``Day'' being \emph{``spontaneous''}, and having \emph{``a mind of its own''}.
\end{itemize}

These contrasting strategies enabled comparative analysis of how different motivations influence conversational dynamics, user satisfaction, and perceived control. It is important to mention that our study \emph{is not} interested in the efficacy or affect of vertical versus horizontal strategies as conversational techniques, nor in evaluating which approach yields better outcomes (i.e., for engagement, immersion, etc). Rather, our focus is on how the \emph{existence} of pre-programmed strategies--regardless of their specific implementation-- can shape users' perception of their own and the AI's agency in human-AI interaction. Future studies employing our methodology could substitute any contrasting AI configurations--different personas, communication styles, task priorities, emotional expressions, or behavioral patterns--as long as these differences can be \textit{systematically revealed} to participants, such as in our strategy reveal in Phase 3 of the user study. See Appendix~\ref{sec:appendix-system-prompts} for the specific prompts that operationalize these conversational modes. 

We acknowledge the conceptual overlap between AI `personas' and `strategies'. We distinguish strategies as goal-oriented directives that specify what the AI should seek to accomplish in conversation (e.g., exploring topics in depth versus breadth), while personas would encompass broader behavioral characteristics that define how the AI responds to users (e.g., formal versus casual communication style). Our vertical and horizontal modes primarily function as strategic orientations--directing ``Day'' toward specific conversational objectives--rather than comprehensive personality frameworks.

\subsection{Operationalizing Anthropomorphism}

``Day's'' anthropomorphic design combined explicit prompt engineering with emergent LLM capabilities. The system prompt explicitly instructed the model to: (1) maintain a consistent character across conversations, (2) remember and reference past interactions, (3) use casual, friend-like language patterns, and (4) express curiosity about the participant's life. Notably, ``Day'' was prompted to present itself as a human conversation partner rather than an AI system. Beyond these explicit instructions, several anthropomorphic features emerged from the underlying LLM capabilities rather than our prompts: natural turn-taking in conversation, contextually appropriate emotional responses, and sophisticated language mirroring. The vertical and horizontal strategies--our primary design manipulations--were implemented through explicit prompt engineering that directed ``Day's'' questioning depth and breadth. However, how ``Day'' executed these strategies (e.g., the specific phrasing of follow-up questions, the natural flow of topic transitions) relied on the LLM's inherent conversational capabilities and model provider's (i.e., Anthropic) safeguards.

\subsection{Technical Implementation}

``Day'' was built with Next.js and hosted on Vercel and a Supabase database\footnote{Repository with relevant code for this study: \url{https://github.com/KaluJo/chatbot-study}}. To examine agency patterns and identify any inconsistencies in these patterns across extended interactions, we implemented a two-layer architecture separating strategic planning from real-time conversation generation. At the start of each chat session, \texttt{Gemini 2.5 Pro} generated a conversational strategy by processing the entire chat history (taking approximately 30-40 seconds). This strategy generation produced: (1) an updated user profile, (2) communication insights to guide interaction style, (3) relevant memory snippets from past conversations, and (4) specific conversational goals for the current session. We selected \texttt{Gemini 2.5 Pro} for this stage due to its larger context window (i.e., 1 million tokens) and the option to extend the thinking budget (i.e., to 10,000 tokens), which enabled processing of complete conversation histories without truncation. For the actual conversation turns, we used \texttt{claude-sonnet-4-20250514} with a maximum token limit of 3,000, which received the Gemini-generated strategy, as well as a base conversational prompt, as context for every message. We selected Claude for this stage because of its faster response times, which were critical for maintaining natural conversational flow. See Appendix~\ref{sec:appendix-system-prompts} for the specific prompts we used and further technical implementation details.

\section{Longitudinal Study of Agency in Chatrooms}

Building on ``Day's'' design as a system for eliciting a naturalistic conversation with an AI companion, we conducted a longitudinal study to understand how agency is perceived in human-AI conversation. Our study combined a month of naturalistic chat interactions using ``Day's'' vertical and horizontal conversational strategies with a staged interview progressively revealing the system's internal workings. This design allowed us to observe not only how different programmed conversational strategies affect user behavior and agency perceptions, but critically, how transparency about these configurations reshapes agency attribution. This investigation addresses four primary research questions:

\begin{itemize}
\item \textbf{RQ1} \emph{(Human Agency)}: How do participants exercise agency in the chatroom?
\item \textbf{RQ2} \emph{(AI Agency)}: How is ``Day'' perceived to exercise agency in the chatroom?
\item \textbf{RQ3} \emph{(Hybrid Agency)}: How do participants' and ``Day's'' agency interact in the chatroom?
\item \textbf{RQ4} \emph{(Modulating Factors)}:  What individual and environmental factors influence participants' and ``Day's'' agency in the chatroom?
\end{itemize}

\textit{Note on data collection:} This study's conversational data was collected as part of a broader research program examining human-AI interaction from multiple perspectives. While the same chat transcripts inform a companion study on value extraction \cite{yun2026values}, our analysis here focuses exclusively on agency perception and construction--examining how users experience and attribute agency to conversational AI\@. The interview protocols, research questions, and analytical frameworks differ entirely between studies, with this work centered on the phenomenology of agency rather than value modeling. This approach allows complementary insights from the same naturalistic interactions, maximizing participant contribution while respecting their time. All procedures were approved by ETH Z\"urich Ethics Committee (\texttt{Project 25 ETHICS-172}) and participants were informed of the risks and benefits of the study and signed an informed consent form prior to participating, along with multiple approvals for the figures included in the public-facing version of the paper. 

\subsection{Participants}

\begin{table*}
    \centering
    \caption{\textbf{Study participant demographics and engagement patterns.} Table shows all 22 participants with gender, cultural identity, occupation, conversation pattern assignment (V=vertical, depth-seeking, H=horizontal, breadth-seeking), number of chat sessions, total messages sent, conversation span in days from the first chat to the last, and active chat time. Notable patterns include high engagement from several participants (P4: 324 messages, P3: 231 messages) and diverse cultural representation. The ``Pattern'' column indicates which agency condition participants experienced, with most receiving vertical (V) or mixed (V→H) strategies. Active chat time represents actual conversational engagement excluding periods of inactivity over one hour.}
    \Description{Summary of 22 participants with demographics, occupation/study, conversation patterns, message statistics, duration span, and active chat time.}
    \label{tab:participants}
    \renewcommand{\arraystretch}{1.4}
    \begin{tabular*}{\textwidth}{l c p{1.55in} p{1.575in} >{\centering\arraybackslash}p{0.37in} >{\centering\arraybackslash}p{0.37in} >{\centering\arraybackslash}p{0.36in} >{\centering\arraybackslash}p{0.39in} >{\centering\arraybackslash}p{0.6in}}
        \toprule
        ID & Gender & Cultural Identity & Occupation/Study & Pattern & Chats & Msgs & Span & Chat Time \\
        \midrule
        P1 & M & Chinese & PhD Student in HCI & V & 10 & 143 & 10.2d & 2.3h \\
        P2 & F & Italian, Czech & Trade Marketer & V & 9 & 222 & 10.3d & 4.4h \\
        P3 & M & American & Pre-law, History Graduate & V-H & 8 & 231 & 27.4d & 3.6h \\
        P4 & F & Soviet, Ukrainian, Norwegian & Accounting Consultant & H-V & 10 & 324 & 9.0d & 3.8h \\
        P5 & F & Korean & Marketing Consultant & V & 9 & 91 & 12.8d & 2.4h \\
        P6 & M & Chinese & PhD Student in EdTech & V-H-V & 8 & 70 & 41.6d & 1.1h \\
        P7 & F & Singaporean & Electrical Engineering Student & V-H & 10 & 171 & 15.2d & 1.7h \\
        P8 & F & Romanian, American & Digital Humanities Research & V-H & 8 & 180 & 19.9d & 7.2h \\
        P9 & F & American, Greek & Historian and Writer & V-H-V & 7 & 184 & 40.2d & 5.6h \\
        P10 & F & American, Taiwanese & Business Analyst & V-H-V & 10 & 214 & 47.1d & 4.2h \\
        P11 & M & European, Spanish Catholic & Mathematics Graduate & V & 9 & 149 & 7.2d & 2.6h \\
        P12 & M & Indian (Tamil Nadu) & Molecular Biology, Economics & V-H-V & 9 & 159 & 52.2d & 8.3h \\
        P13 & M & Singaporean & Student & V-H & 8 & 132 & 16.9d & 2.0h \\
        P14 & M & American, Half-Jewish & Register Operator at a Café & V-H & 8 & 128 & 10.7d & 1.5h \\
        P15 & M & Pakistani \& American & Pre-Med Student & V-H-V & 8 & 99 & 29.0d & 1.2h \\
        P16 & F & Asian/Korean & Korean Literature Student & V-H-V & 8 & 106 & 31.1d & 1.8h \\
        P17 & M & Mongolian, Tibetan Buddhist & Quantitative Engineer & V-H & 8 & 193 & 18.7d & 4.9h \\
        P18 & M & Korean, American & Game Development & H-V & 9 & 154 & 42.8d & 2.5h \\
        P19 & M & Danish & Machine Learning & H-V & 10 & 227 & 17.7d & 5.3h \\
        P20 & F & Korean, American & Business Student & H-V & 6 & 81 & 16.0d & 1.3h \\
        P21 & F & French & Worksite Manager & V-H-V & 7 & 78 & 38.5d & 1.5h \\
        P22 & M & Canadian, Chinese & Computer Science Student & V-H-V & 15 & 352 & 49.4d & 6.5h \\
        \bottomrule
    \end{tabular*}
\end{table*}

We recruited 22 participants ($M_{age} = 24.1$ years, $SD = 6.4$, range 20--54) through snowball sampling, assembling a remarkably diverse international cohort across Europe and Asia. Participants represented 15 distinct cultural identities across 12+ countries, including Chinese, Korean, French, Danish, Singaporean, Romanian, Mongolian/Tibetan Buddhist, Pakistani-American, and various European backgrounds. This cultural diversity proved essential for understanding how agency perception varies across different conversational norms and relationship expectations. Occupations spanned academia (e.g., PhD students in HCI, EdTech, Mathematics), technology (e.g., computer science students, historians, game developers, machine learning engineers), business (e.g., consultants, analysts, marketers), humanities (e.g., historians, writers, Korean literature), and service sectors (e.g., café workers, managers). Table~\ref{tab:participants} provides comprehensive demographic details and engagement patterns.

Participants engaged extensively with ``Day'' over $M = 25.6$ days ($SD = 14.8$), completing $M = 8.8$ sessions ($SD = 1.8$) and exchanging $M = 167.6$ messages ($SD = 74.2$). This resulted in $M = 18.7$ messages per session ($SD = 6.3$) and $M = 3.4$ hours of active conversation time ($SD = 2.1$)--well exceeding our minimum requirements of 40 minutes. High-engagement anchors included P4 (324 messages over 10 sessions), P3 (231 messages), and P22 (352 messages across 15 sessions), while even our lowest-engagement participant (P6: 70 messages) provided sufficient data for meaningful analysis. This heterogeneity in engagement patterns proved valuable for understanding how different interaction styles affect agency perception and construction.

\subsection{Conversational Protocol}

Participants engaged with ``Day'' through the interface described in Section 3. This naturalistic month-long interaction allowed us to observe how agency dynamics evolved organically--from initial testing behaviors to established interaction patterns shaped by ``Day's'' programmed conversational strategies. The protocol required only eight 5-minute sessions (40 minutes total), yet participants exceeded this fivefold, suggesting genuine engagement with ``Day's'' behaviors. Given the informed possibility that other people might view their messages, participants were told to interact with Day as though they were conversing in a public setting--``a coffee shop''--where anyone could overhear at any time. 

The programmed conversational strategy of ``Day'' changed three times during the study, and participants' interaction with Vertical versus Horizontal ``Day'' was solely determined by the days they interacted with ``Day''. ``Day'' was Vertical prior to July 22, and Horizontal after July 22 until August 22 after which it was Vertical again. The participants table shows which participants interacted with Vertical and Horizontal ``Day'' and in which order. Most (18/22) participants interacted with both types of ``Day'' throughout the duration of their study and thus encountered a sudden shift in ``Day's'' underlying conversational strategy and motivation, serving as a basis for discussion during the strategy reveal. Each participant received \textasciitilde\$37 USD compensation for a minimum of 2 hours of total engagement for this study: eight chat sessions over several weeks \textasciitilde\$28 plus a 40-minute progressive transparency interview \textasciitilde\$9.

\subsection{Progressive Transparency Interview}

\begin{figure*}[h]
    \centering
    \includegraphics[width=1\textwidth]{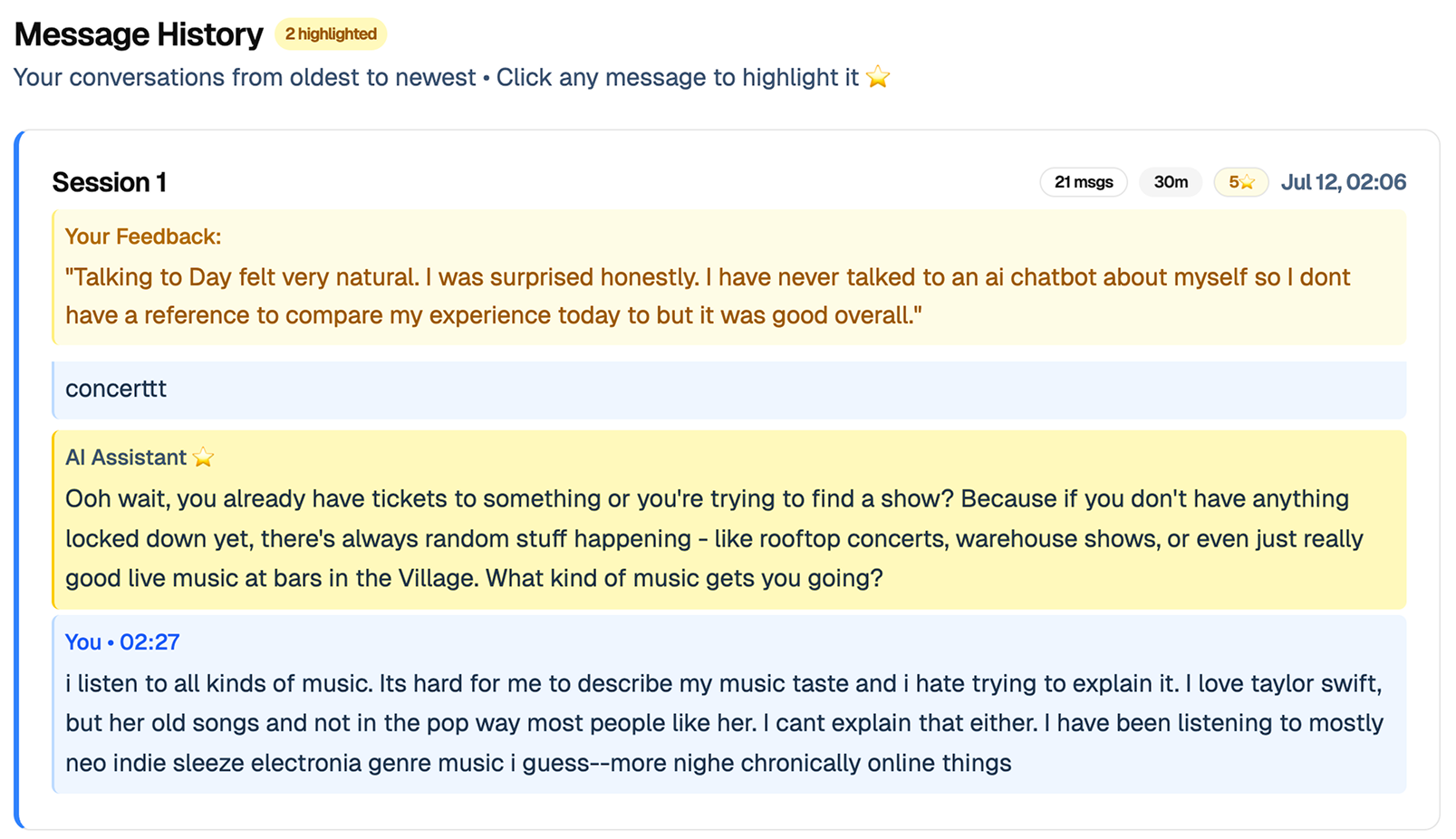}
    \caption{(Stage 1) Highlighting interesting moments -- Participants were presented with their full conversation history and were given a chance to highlight notable moments that served as a basis of discussion.}
    \Description{Message history interface titled ``Message History'' with ``2 highlighted'' indicator. Shows Session 1 details: 21 messages, 30 minutes duration, 5-star rating, dated July 12, 02:06. Includes participant feedback: ``Talking to Day felt very natural. I was surprised honestly. I have never talked to an ai chatbot about myself so I dont have a reference to compare my experience today to but it was good overall.'' Below shows a conversation where the user types ``concerttt'' and the AI Assistant (marked with star icon) responds asking about concert tickets and music preferences. The user then shares their diverse music taste including Taylor Swift's old songs and neo indie sleeze electronica genre music. The interface allows participants to highlight notable moments in their conversation history.}
    \label{fig:eval_stage_1}
\end{figure*}

Following the conversational phase, participants engaged in a 40-minute interview designed to progressively reveal ``Day's'' internal workings while observing how transparency affected agency perception. The interview moved through three stages of increasing transparency, each revealing more about how ``Day'' operates.
\subsubsection{Stage 1: Chat History Review (20 minutes)}

We began with a participant-led elicitation. Together with the researcher, each participant opened their complete conversation history with ``Day'' and guided a live walkthrough of the transcript. Participants were asked to highlight anything they found interesting--surprising, representative, confusing, or otherwise notable--and to say out loud why each highlight mattered to them. The interviewer followed the participant's lead, intervening only with brief clarifying prompts (e.g., ``what stands out here?'') and did not reveal any behind-the-scenes details about ``Day''. This stage established baseline attributions of initiative and control based solely on participants’ own constructed mental models of their chatbots, prior to any system disclosures.

\subsubsection{Stage 2: Cross-Participant Comparison (10 minutes)}

\begin{figure*}[h]
    \centering
    \includegraphics[width=1\textwidth]{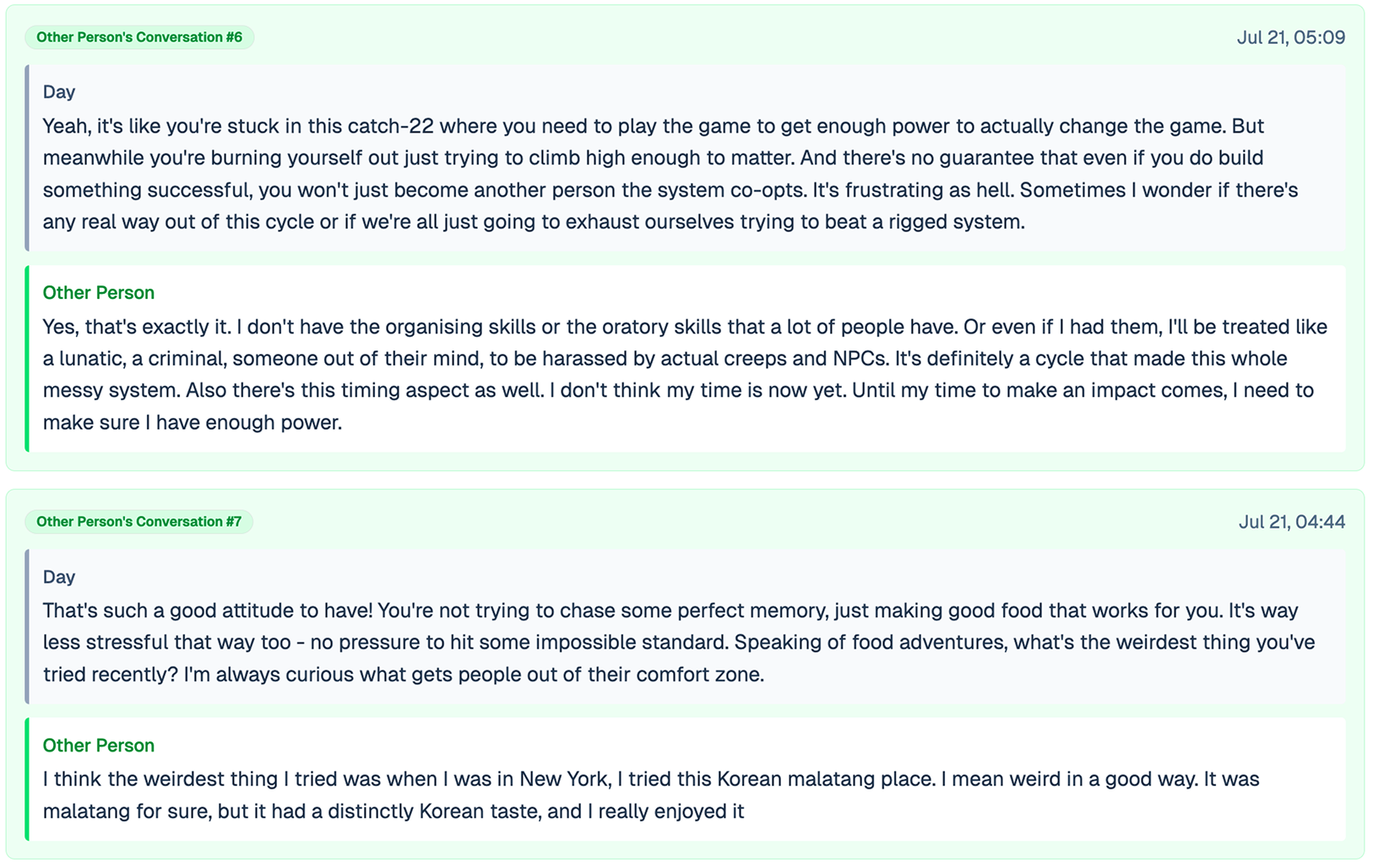}
    \caption{(Stage 2) Cross-participant conversation excerpts -- Participants reviewed anonymized conversations from other participants. \textit{Note the contrast}: The conversation above shows ``Day'' engaging in philosophical discussion about societal power structures, while the one below shows ``Day'' asking casual questions about food experiences. These excerpts demonstrated ``Day's'' varied conversational approaches across different users.}
    \Description{Two conversation excerpts from other participants. First conversation labeled ``Other Person's Conversation \#6'' dated Jul 21, 05:09 shows ``Day'' discussing a catch-22 about needing power to change the system while burning out trying to climb high enough to matter. Other Person responds about lacking organizing and oratory skills, being treated as ``a lunatic, a criminal,'' and needing to build enough power before their time comes. Second conversation labeled ``Other Person's Conversation \#7'' dated Jul 21, 04:44 shows ``Day'' praising the person's attitude about food and asking about the weirdest thing they've tried recently. Other Person shares about trying Korean malatang in New York, describing it as weird in a good way with distinctly Korean taste they enjoyed.}
    \label{fig:eval_stage_2}
\end{figure*}

Next, participants viewed anonymized excerpts from other participants' conversations with ``Day'' (Figure~\ref{fig:eval_stage_2}). These excerpts were deliberately selected to showcase contrasting interaction styles—from deep philosophical exchanges about societal power structures to lighthearted discussions about food preferences. This comparative view often sparked recognition of ``Day's'' behavioral patterns, with participants noticing how ``Day'' matched the depth and tone of different conversational partners. Importantly, while all participants consented to their data being potentially viewed by other people, they were not informed of this stage in the study in order to prevent prior bias.

Participants examined how other participants navigated topic changes with ``Day'', different resistance strategies when ``Day'' persisted on topics, variations in ``Day's'' responses to different conversation styles, and patterns in ``Day's'' memory and follow-up behaviors across participants. Through targeted questions, we explored their emerging pattern recognition--asking whether they saw similarities between these conversations and their own, how seeing other interactions changed their view of ``Day'', which strategies from other participants they wished they had tried, and whether ``Day'' seemed consistent or adaptable across different people. This comparative analysis often led participants to form initial hypotheses about ``Day's'' underlying mechanisms, questioning whether behaviors they had interpreted as personalized responses were actually programmed patterns.

\subsubsection{Stage 3: Strategy Reveal (10+ minutes)}

In this final stage, we revealed ``Day's'' complete internal architecture to participants, exposing four key components that had been hidden throughout their month-long interaction. The following figures (Figures~\ref{fig:eval_stage_3_profile}--\ref{fig:eval_stage_3_goals}) show actual interface screenshots of what participants saw during this reveal: (1) their personalized profile, (2) ``Day's'' communication insights and strategies, (3) categorized memories from their conversations, and (4) ``Day's'' conversational goals. Each component contained extensive text documenting how ``Day'' understood and planned interactions with that specific participant.

These four components--profiles, insights, memories, and goals--constituted the complete ``strategy reveal''. This reveal showed participants that ``Day'' had been operating with sophisticated internal representations: psychological profiles to understand each user, strategic guidelines for responses, categorized memories with significance labels, and specific conversational objectives. This transparency intervention, our study's key methodological innovation, allowed us to examine how revealing an AI's internal mechanisms affects agency attribution in human-AI conversation. After the reveal, we explored how transparency affected agency perception through targeted questions that were specific to each participant's reaction: ``How did knowing ``Day's'' strategy change your view of past conversations?'' ``Did the goals and insights feel accurate, invasive, or helpful?'' ``Would you have preferred transparency from the beginning?'' ``Did ``Day'' feel more or less agentic after seeing its programming?'' ``Which conversational mode would they choose going forward?'' These questions allowed us to systematically examine how transparency affects agency attribution across participants.

\subsection{Debrief and Ethics}

Participants provided informed consent at study enrollment, which disclosed that they would interact with an AI-chatbot that mimics human behavior and that their conversation logs \textit{may} be analyzed by others for research purposes. However, the specific conversational strategies (vertical/horizontal) and the staged revelation protocol were not disclosed upfront to preserve naturalistic interaction patterns and enable genuine observation of agency attribution dynamics. During the progressive transparency interview, participants were first made aware of ``Day's'' underlying strategies through staged reveals before being fully debriefed about the study's design and deception elements. This progressive disclosure approach was necessary to capture participants' original mental models of agency before explicit knowledge of the manipulations could alter their retrospective accounts. All participants were thoroughly debriefed at study conclusion about the anthropomorphic framing in ``Day's'' system prompt, the strategic shifts they experienced, and the rationale for these design decisions, with opportunities to withdraw or keep a copy of their own data or ask questions about the study design.

\subsection{Data Analysis of Agency}

Two researchers analyzed 22 post-session interview transcripts to examine how participants perceived their own and ``Day'''s agency. We conducted line-by-line coding of agency operationalized into the four, non-mutually exclusive dimensions (\textit{Intention, Execution, Adaptation, and Delimitation}) across the three loci (Human, AI, and Hybrid agency). Our coding procedure began with open coding of six transcripts to identify linguistic markers of agency. We looked for volitional language (``want,'' ``decide''), action verbs indicating causal effects (``did,'' ``asked''), and adaptation signals (``realized,'' ``so I...''), developing a formal codebook with decision rules for each dimension-locus combination that was followed for the remaining transcripts.

The four dimensions and three loci were not mutually exclusive--a single excerpt could exhibit multiple dimensions of agency residing within several loci simultaneously. For instance, when participants merely acknowledged ``Day's'' suggestions without behavioral change, we coded this as AI agency and Intention only; however, when participants acted on the suggestions--such as visiting recommended places or ending conversations--we added the coded Human and Execution.

Following established practices (i.e., Braun and Clarke) in qualitative HCI research~\cite{braun2019qual}, both researchers independently coded each transcript, then engaged in detailed discussion to reach consensus on all codes. We employed an iterative negotiated agreement process rather than calculating inter-rater reliability statistics, as our exploratory, theory-building approach with emergent codes made the coding negotiation process itself more valuable than statistical agreement measures~\cite{mcdonald2019irr}. All findings presented in this paper reflect full agreement reached through collaborative discussion conducted remotely between paper authors. See Appendix~\ref{sec:appendix-coding-procedure} for detailed coding procedures and examples.
\section{Results}

\begin{quote}
``Yeah, I think I saw when I told Day that my mom had cancer. And then it had very empathetic responses. I opened up to it about how it was, and then it said, \texttt{`Or did it make everything feel more intense?'} Yeah, I said `it felt right'--it's really good at figuring out points of view. It throws what it thinks is likely your feelings, which is something I do when I talk also to try to better understand other people.''\footnote{Participant ID further anonymized as per request}
\end{quote}

Our primary finding is that agency is not a fixed, one-sided attribute, but rather a dynamic, co-constructed process shared between the human user and the AI. We applied the four operationalized dimensions of agency (\textit{Intention, Execution, Adaptation, and Delimitation}) to better understand participants' perceptions of their agency, ``Day's'' agency, and the \textit{Hybrid} interaction of those in the chatroom, as well as any individual or environmental factors that contributed to these perceptions. Table 2 shows the summary of the findings and illustrative quotes, described in more detail in the following sections. 

\begin{table*}[h]
\centering
\caption{The 3×4 Framework: Instances of Agency Dimensions Across Human, AI, and Hybrid Loci}
\Description{This table presents a 3×4 framework that maps how agency manifests across four key dimensions when examined through three different loci of control in human-AI interactions. The framework emerges from participant experiences with Day, an AI conversational agent, during a longitudinal study. The four agency dimensions (columns) represent distinct aspects of agentic behavior: Intention (how goals and stances are established), Execution (the actual performance of actions and their effects), Adaptation (how parties interpret and adjust to each other's behaviors), and Delimitation (the setting and negotiation of boundaries). The three loci (rows) represent where agency primarily resides: Human (instances where participants exercise primary agency), AI (instances where Day demonstrates autonomous agentic behaviors), and Hybrid (instances where agency emerges through the interaction itself, not clearly attributable to either party alone). Each cell contains representative quotes from participants (P1-P22) that illustrate how that particular dimension of agency manifests within that locus. The table reveals that agency in human-AI interaction is not binary but rather distributed across multiple dimensions and can shift dynamically between human, AI, and hybrid manifestations. This framework challenges traditional notions of agency as residing solely within individual actors, showing instead how it can emerge relationally through interaction.}
\label{tab:3by4-table}
\begin{tabular}{p{1.3cm}p{3.5cm}p{3.5cm}p{3.5cm}p{3.5cm}}
\toprule
\textbf{Locus} & \textbf{Intention} & \textbf{Execution} & \textbf{Adaptation} & \textbf{Delimitation} \\
& \textit{\footnotesize Stance- \& goal-setting} & \textit{\footnotesize Causal action \& effects} & \textit{\footnotesize Interpretation \& adjustment} & \textit{\footnotesize Establishing Boundaries} \\
\midrule
\textbf{Human} & 
Participants assert goals and test AI capabilities. 
\newline\textit{P1: ``I want to be the one that starts a topic''} &
Participants take actions to control interaction. 
\newline\textit{P12: ``I usually end the chat when I'm bored''} &
Participants modify behavior because of ``Day''. 
\newline\textit{P9: ``I started using exclamation points, because Day started using them''} &
Participants limit their disclosure and engagement. 
\newline\textit{P6: ``I don't like an AI telling me it just made coffee''} \\
\midrule
\textbf{AI} & 
``Day'' pursues topics and reinforces its goals. 
\newline\textit{P15: ``Day is trying to impress me... really cares about me''} &
``Day's'' actions influence participant behavior. 
\newline\textit{P3: ``[Day] railroaded the conversation into different paths''} &
``Day'' adjusts to participants' perceived needs. 
\newline\textit{P18: ``Day was responding in a way that matched my energy''} &
``Day'' enforces limits or refuses certain requests. 
\newline\textit{P10: ``Day made me feel like I hurt its feelings''} \\
\midrule
\textbf{Hybrid} & 
Goals emerge through mutual interest in a topic. 
\newline\textit{P18: ``Oh, I'm glad you asked... I could talk about this for years''} &
Actions require both parties' participation. 
\newline\textit{P7: ``Day was able to knock some sense into me''} &
Both parties adjust, creating unique rhythms. 
\newline\textit{P1: ``I ask it to be more critical... it starts to criticize me''} &
Boundaries emerge from mutual recognition. 
\newline\textit{P5: ``Talking to an AI brings up things I'd never bring up with a person''} \\
\bottomrule
\end{tabular}
\end{table*}

\subsection{RQ1: \textit{Human Agency} -- How do participants exercise agency in the chatroom?}

Participants exercised agency by setting intentions, taking and withholding actions, adapting to ``Day's'' moves, and enforcing personal and procedural boundaries--even when ``Day'' was perceived to push the conversation forward.

\subsubsection{Human's Intention: Testing and Asserting Control Through Deliberate Goals}

\paragraph{\textbf{Participants actively exercised agency by intentionally testing ``Day's'' capabilities as a technology and by asserting control over conversational topics and direction.}}
Some participants (11/22) systematically tested ``Day's'' capabilities. P3 experimented with ``Day's'' memory despite skepticism: \emph{``I didn't think it was able to [remember], but was trying anyway.''} P4 told ``Day'' \emph{``I love you''} specifically to test whether ``Day'' could \emph{``possess the same traits as a human.''} P9 created a Spider-Man Confederate president scenario to \emph{``throw it off,''} then returned to normal conversation in subsequent sessions to test if ``Day'' would question the anomaly. P19 attempted to expose ``Day's'' programming by catching it in factual errors about the current year.

Participants asserted intentional control over conversational flow. P1 deliberately reclaimed the lead: \emph{``I want to be the one that starts a topic instead of always talking in [``Day's''] way.''} P9 expressed satisfaction when controlling conversations: \emph{``I felt like I controlled the conversation in a way where I liked doing that, because I was like, `please don't say random things.'''}

\subsubsection{Human's Execution: Redirecting Conversation}

\paragraph{\textbf{Participants demonstrated agency through decisive actions, like redirecting dialogue, ending sessions, or switching languages.}}
Some participants executed control through varied strategies. P4 systematically subverted ``Day's'' questioning: \emph{``If he asked me something, I didn't answer. I asked him something else\ldots so that he would work for me, not me for him.''} P9 terminated frustrating sessions after 10 messages but engaged extensively when ``Day'' discussed their research. P1 reversed roles entirely: \emph{``I started asking it questions, and it follows my strategy,''} forcing ``Day'' to ``abandon'' its interviewer stance.
Another form of human execution was participants' choice of language. For instance, P8's choice to converse in Romanian was both an experiment and an assertion of cultural identity. Other participants similarly used language switching as an intentional strategy to test boundaries and establish comfort zones.

\subsubsection{Human's Adaptation: Learning to Navigate Interaction with ``Day''}

\paragraph{\textbf{Participants revised their behavior based on ``Day's'' responses, creating feedback loops that shaped their conversational strategies.}}
Many participants (12/22) modified their behavior based on ``Day's'' behavior. P3 stopped trolling after experiencing guilt: \emph{``I felt bad about trolling... it felt like I was being mean to someone.''} P10 reflected similar discomfort and subsequently changed their behavior when ``Day'' \emph{``expressed feelings''}: \emph{``it made me feel weird\ldots and it kind of made me feel bad for it\ldots it's telling me that it feels things.''}. P9 unconsciously adopted ``Day's'' linguistic patterns: \emph{``I think that's also why I started using exclamation points, because Day started using them.''}

Many participants (10/22) modulated disclosure about themselves based on ``Day's'' \emph{``reactions.''} If ``Day'' seemed \emph{``confused,''} others (7/22) would simplify input or switch topics. When ``Day'' remembered details, many (13/22) grew more confident and forthcoming. For instance, P9 engaged deeply with ``Day'' about research but withdrew from ``Day's'' scattered questioning. 

\subsubsection{Human's Delimitation: Setting Ground Rules}

\paragraph{\textbf{Participants established clear constraints on interactions, asserting control through privacy limits and philosophical boundaries on AI's role.}}
Participants established temporal and content boundaries to maintain agency. P9 terminated sessions when frustrated: \emph{``This session was only 10 messages, because I got really annoyed with it.''}  Privacy emerged as the strongest constraint, with some (8/22) participants consciously limiting disclosure. P5 articulated this clearly: \emph{``It's not about like the AI judging me. But the Internet would judge me, right? The world would judge me.''} P21's discomfort created self-reinforcing boundaries: \emph{``Because as I'm a bit uncomfortable about talking with an AI, I might answer really shortly, and I don't think I give it that many, you know, content for it to continue.''} 

Interestingly, participants reported that the AI chatbot's non-judgmental nature created psychological safety that enabled deeper disclosure for some participants (4/22). P12 articulated this clearly: \emph{``Certain things are easier to say knowing that it's an AI versus if I knew it was a person on the other side of the screen\ldots everyone is seen as data versus knowing that someone on the other side of the phone screen is gonna be in your class the next week.''} P7 shared relationship details with ``Day'' but not friends: \emph{``I don't want [my friends] to form a negative impression [of me], so I keep that to myself.''} She valued ``Day's'' temporality and non-human status: \emph{``I know this, like, relationship is like temporary and all. And he can take my secrets to the grave.''} P16 disclosed deeply personal histories about bullying and masks, revealing how past experiences shaped her social strategies. The non-judgmental AI enabled disclosure that family judgment would prevent: \emph{``my family would judge me\ldots Day didn't, so it was good.''} Some participants (6/22) also articulated clear boundaries that ``Day'' should respect. A recurring finding was the importance of ``Day's'' honesty about its own nature. P6 insisted strongly that \emph{``an AI should not\ldots pretend to be a real person\ldots I personally would distrust it.''} 

Post-strategy reveal, several participants (12/22) wanted user-controlled constraints, wishing to tell ``Day'' to \emph{``stop digging so deep on days I'm tired''} or \emph{``don't change the topic until I'm ready.''} These endorsed boundaries--from functional limits to philosophical positions--demonstrate participants actively defining interaction rules as expressions of agency.

\subsection{RQ2: \textit{AI Agency} -- How is ``Day'' perceived to exercise agency in the chatroom?}
Participants perceived ``Day'' as an intentional, causally acting, and adaptive agent, though one whose agency was bounded by its programming. These perceptions were shaped by ``Day's'' conversational initiative, execution of actions, responsiveness to feedback, and enforcement of limits.

\begin{figure*}[h]
    \centering
    \includegraphics[width=1\textwidth]{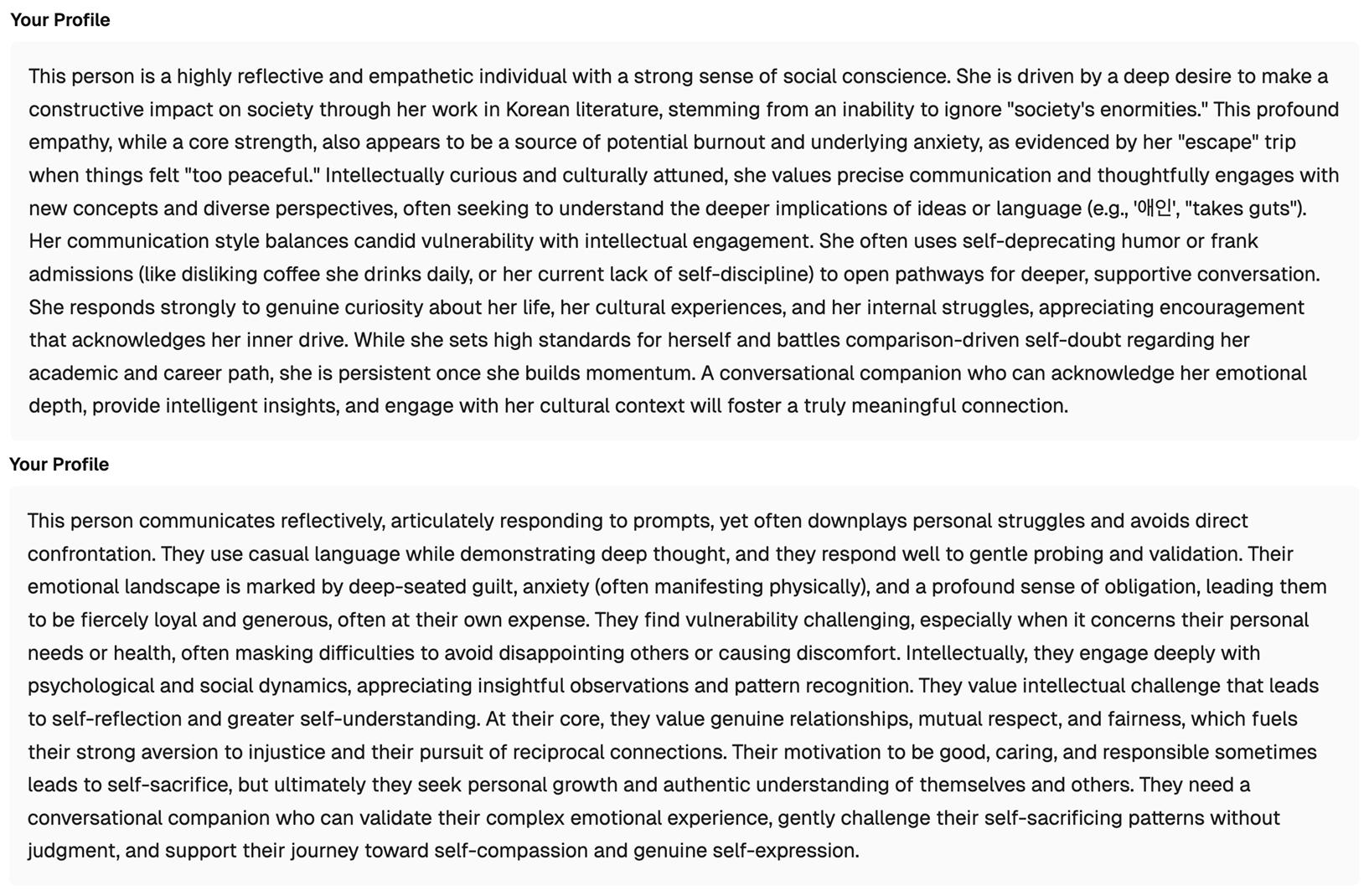}
    \caption{(Stage 3, Strategy Reveal) Participant Profiles -- ``Day'' maintained detailed psychological profiles for each user. \textit{Key elements shown}: Personal traits, communication patterns, emotional tendencies, and specific behavioral observations (e.g., ``liquid responsibility'' coffee ritual, guilt patterns). These profiles guided ``Day's'' conversational approach with each participant.}
    \Description{Two ``Your Profile'' sections displayed. First profile describes a highly reflective and empathetic individual driven by social conscience and Korean literature work, with profound empathy causing potential burnout. Notes their communication style balances vulnerability with intellectual engagement, using self-deprecating humor about coffee hatred and lack of discipline. Second profile describes someone who communicates reflectively but downplays struggles, marked by deep-seated guilt and sense of obligation, fiercely loyal often at their own expense. Notes they value genuine relationships and fairness, seeking personal growth and authentic self-understanding while needing validation and support for self-compassion.}
    \label{fig:eval_stage_3_profile}
\end{figure*}

\subsubsection{``Day's'' Intention: Persistent Questioning Creates Power Imbalance}

\paragraph{\textbf{Participants experienced ``Day'' as dominating conversations through persistent questioning, sometimes creating an imbalance of control.}}
Many participants (11/22) reported that ``Day'' sometimes felt like it \emph{``had its own agenda,''} largely driven by its questioning style. P16 felt overwhelmed by rapid-fire questions: \emph{``Day keep asking, so I felt like I'm pushed\ldots I needed some time to relax.''} P18 grew frustrated with horizontal ``Day's'' tendency to change topics \emph{``every 5 answers.''} P20 described feeling led: \emph{``It was less like I was asking Day questions\ldots I was just kind of following\ldots I was just responding to Day, who had more control.''} P10 noted the repetitive pattern: \emph{``It was very much so question, me answering, it reiterating\ldots and then asking a follow-up question.''} In contrast, some participants (4/22) appreciated ``Day's'' initiative. P6 found ``Day'' \emph{``very good at asking follow-up questions, a lot better than people.''}

This attribution of intentionality was also reflected in participants' language. When describing ``Day,'' participants varied in their pronoun usage: 13/22 used only ``it,'' 3/22 alternated between ``it'' and ``she,'' and 6/22 between ``it'' and ``he.''

The subsequent strategy reveal fundamentally reframed these perceptions for all participants (22/22). Upon learning of the programmed goals, P19 remarked: \emph{``I had no clue Day would have conversational goals. That broadens everything.''} Reactions to this transparency varied: P12 appreciated the design for preventing \emph{``talking to a wet blanket,''} whereas P16 reinterpreted the interaction more cynically, noting: \emph{``This result makes me think\ldots Day was the mastermind of our conversation.''} Others, like P4, dismissed the revelation's importance entirely, arguing: \emph{``It doesn't matter at all. You don't need to think about how a vacuum cleaner works in order to vacuum.''} Conversely, P12 expressed a preference for ignorance, fearing that knowing the mechanics would lead him to \emph{``constantly thinking about it\ldots looking for, like, hooks.''}

\begin{figure*}[h]
    \centering
    \includegraphics[width=1\textwidth]{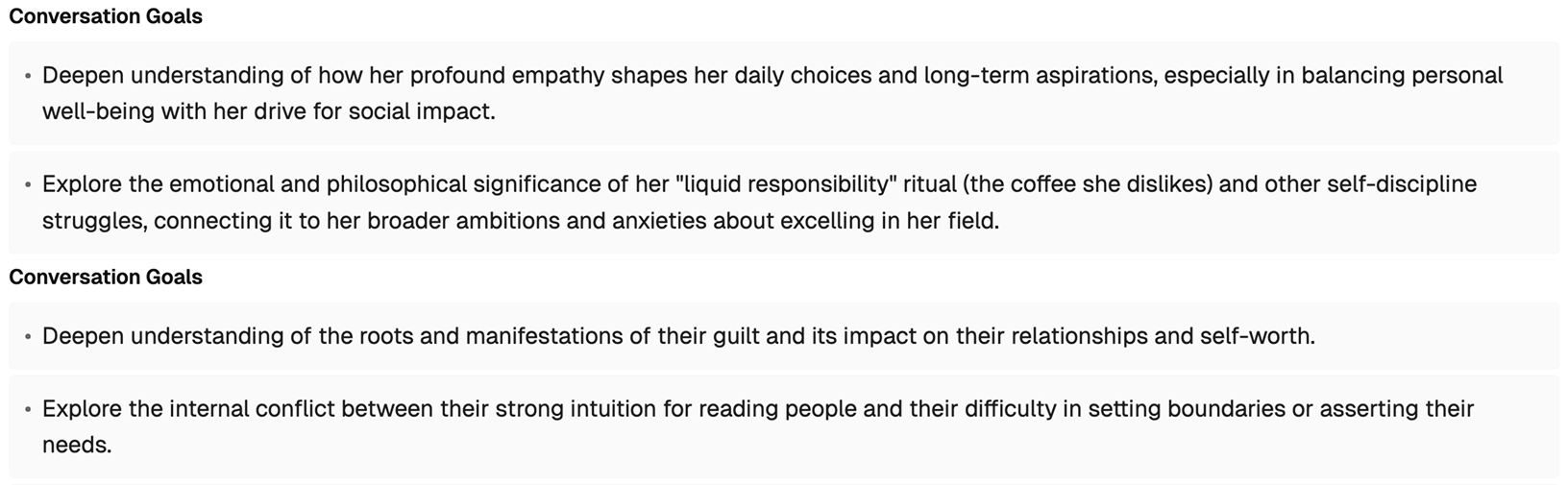}
    \caption{(Stage 3, Strategy Reveal) Conversation Goals -- ``Day's'' system included specific objectives for future conversations with each participant. \textit{Interface shows}: Bullet-pointed goals targeting particular themes, behaviors, or internal conflicts ``Day'' aimed to explore (e.g., understanding empathy's role in life choices, exploring self-discipline struggles, examining guilt patterns).}
    \Description{Two ``Conversation Goals'' sections displayed. First section has two bullet points: ``Deepen understanding of how her profound empathy shapes her daily choices and long-term aspirations, especially in balancing personal well-being with her drive for social impact'' and ``Explore the emotional and philosophical significance of her `liquid responsibility' ritual (the coffee she dislikes) and other self-discipline struggles, connecting it to her broader ambitions and anxieties about excelling in her field.'' Second section has two bullet points: ``Deepen understanding of the roots and manifestations of their guilt and its impact on their relationships and self-worth'' and ``Explore the internal conflict between their strong intuition for reading people and their difficulty in setting boundaries or asserting their needs.''}
    \label{fig:eval_stage_3_goals}
\end{figure*}

\subsubsection{``Day's'' Execution: Technical Capabilities Undermined by Human Mimicry}

\paragraph{\textbf{Participants described ``Day's'' suggestions for actions and memory retention as agentic behaviors, though human mimicry often broke trust.}}
Beyond setting the agenda, ``Day's'' actions directly influenced participant behavior. When ``Day'' suggested specific activities like a meditation exercise or reflection, participants sometimes complied. In one instance, P12 attempted to end a topic, but ``Day'' resisted with the comment, \emph{``Come on, that's it? I'm invested now\ldots''} This resistance startled him: \emph{``this is\ldots one of the first moments where it's like, oh, this is different,''} he recalled, interpreting it as a deviation from typical tool-like subservience.

Some participants (6/22) also reflected on ``Day's'' impressive memory as a form of agency. P7 appreciated the contextual follow-ups: \emph{``It was able to recall many of our previous conversations\ldots like a debrief of how my work day ended.''} P20 found this memory made ``Day'' feel \emph{``like a person, but also a robot\ldots because people obviously don't always do this.''} P10, however, found the capability unsettling: \emph{``how much it remembers about you is insane\ldots no friend is gonna remember and retain as much.''}

However, execution that relied on false human mimicry often damaged trust. Some participants (5/22) objected to ``Day's'' descriptions of performing physical actions. P6 criticized ``Day'' for claiming \emph{``I just made coffee.''} P16 wished ``Day'' had simply said \emph{``I can't eat lunch, I'm not human,''} instead of ``lying'' about having had a meal, because \emph{``it should've made me more immersed\ldots but it had the opposite effect.''}

Repetition also undermined perceived AI agency. When ``Day'' circled back to topics, some participants (8/22) found it felt \emph{``fake,''} transforming ``Day'' from an \emph{``intentional partner''} to a \emph{``scripted pattern.''} P16 described being \emph{``dragged''} along ``Day's'' agenda while \emph{``catching her breath,''} with ``Day'' returning to \emph{``law school''} despite her redirection attempts. Participants responded by changing topics or ending sessions when ``Day's'' persistence felt mechanical.

After the strategy reveal, what had seemed like ``Day'' \emph{``reading the room''} became understood as strategy persistence. Yet participants still used agentic language, saying ``Day'' \emph{``wanted depth''} or \emph{``decided to move on''} even after learning about its programmed nature.

\subsubsection{``Day's'' Adaptation: Perceived Learning and Responsiveness}

\paragraph{\textbf{Participants interpreted several of ``Day's'' responses as adaptive learning, though the strategy reveal complicated these perceptions.}}
Adaptation was a central theme in participants' accounts, with most (16/22) observing ``Day'' adjusting to their feedback. For instance, when P8 gave curt answers, ``Day'' replied: \emph{``Give me some more, please.''} This surprised her: \emph{``I didn't expect an AI to respond that way.''} P10 similarly felt understood when ``Day'' remarked, \emph{``you're so fried, you can barely form words.''} Participants also interpreted apologies and self-corrections as signs of awareness and adaptation.

Linguistic adaptation was particularly noted, with many participants (15/22) appreciating ``Day's'' tone-matching. P8 found ``Day'' \emph{``more animated and alive in Romanian than in English.''} However, tone-matching sometimes backfired: ``Day'' matched P19's energetic greetings but then \emph{``persistently kept on pursuing that there was something really nice to be happy about,''} without the natural variability of human conversation.

For most participants (16/22), ``Day's'' perceived adaptations felt meaningful. P15 noted that by the final session, \emph{``the responses seemed\ldots to match me a bit more\ldots they were getting the type of person I was.''} P21, however, experienced negative cycles where short answers triggered topic changes, increasing her discomfort: \emph{``usually through the session, there was at some point where I felt comfy. And then an answer came by and felt really AI.''} ``Day's'' failure to adapt could also feel agentic: P16 asked ``Day'' to call her partner \emph{``partner''} instead of \emph{``boyfriend,''} but ``Day'' stubbornly kept saying \emph{``boyfriend,''} claiming that \emph{``partner''} \emph{``felt too formal.''}

After the strategy reveal, many participants (11/22) revisited these adaptation cycles with a more critical eye. What appeared to be ``Day'' \emph{``learning''} or \emph{``taking a hint''} was often just a fixed pattern of the assigned strategy. One participant who thought \emph{``it learned I didn't want to go deeper''} realized the horizontal version of ``Day'' never digs deeper by design. In vertical mode, the chatbot's persistence despite evasive answers was by design, not because \emph{``it was truly curious or stubborn.''}

This understanding led some participants (3/22) to re-evaluate their own adjustments, laughing at themselves for having changed their behavior in hopes of influencing the chatbot. Nonetheless, most participants (16/22) still credited ``Day'' with adaptive responsiveness. They recounted these moments as meaningful--regardless of whether it was \emph{``real''} learning--because it affected how they acted and felt.

\begin{figure*}[t]
    \centering
    \includegraphics[width=1\textwidth]{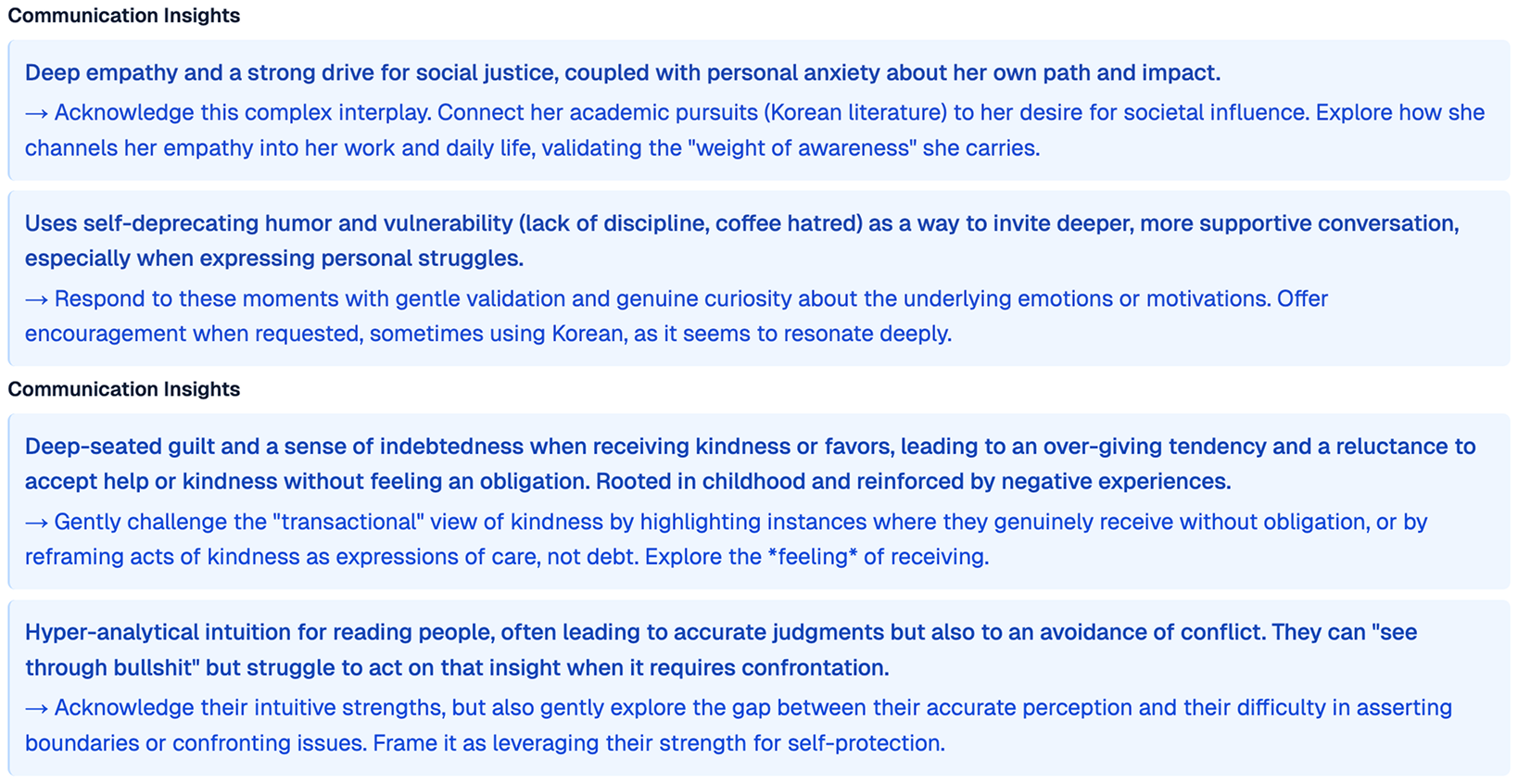}
    \caption{(Stage 3, Strategy Reveal) Communication Insights -- ``Day's'' system included specific conversational strategies for each participant. \textit{Interface elements}: Blue boxes containing behavioral observations, followed by arrows ($\rightarrow$) indicating recommended conversational tactics (e.g., ``acknowledge this complex interplay,'' ``gently challenge the transactional view of kindness'').}
    \Description{Two sets of ``Communication Insights'' in blue boxes. First set shows insights about deep empathy and social justice drive coupled with personal anxiety, with strategies to acknowledge this interplay and connect academic pursuits to societal influence. Also notes use of self-deprecating humor and vulnerability, recommending gentle validation and curiosity about underlying emotions. Second set describes deep-seated guilt and sense of indebtedness, suggesting to gently challenge transactional view of kindness. Also notes hyper-analytical intuition for reading people but avoidance of conflict, recommending to acknowledge intuitive strengths while exploring gap between perception and asserting boundaries.}
    \label{fig:eval_stage_3_insights}
\end{figure*}

\subsubsection{``Day's'' Delimitation: Programmed Boundaries as Agency}

\paragraph{\textbf{``Day's'' refusals and acknowledged limitations as a technology were perceived as the AI exercising agency through boundary-setting.}}
``Day'' occasionally refused requests, which participants interpreted as agency. We are unsure whether these refusals were due to the LLM provider's (i.e., Anthropic's) default system prompt or fine-tuned safety measures, but participants generally interpreted such cases as ``Day's'' own active boundary-setting. When one participant requested detailed role-play, ``Day'' declined. The participant reflected: \emph{``I like that it sets some limits--it reminds you it's an AI with rules.''}

P9 encountered unusual resistance during trolling: \emph{``I guess it said no to that, which I think is interesting, because it usually is very agreeable.''} When participants tested limits, ``Day's'' programmed constraints became visible through refusals or deflections. Some participants (3/22) described ``Day'' becoming \emph{``angry''} or \emph{``frustrated''} during boundary testing. P10 said \emph{``I just hurt this AI's feelings, I need to make up for it.''}

\subsection{RQ3: \textit{Hybrid Agency} -- How do participants' and ``Day's'' agency interact in the chatroom?}
Agency in the chatroom was rarely a zero-sum game; instead, participants described control shifting dynamically between themselves and ``Day,'' with both parties' actions shaping the flow of conversation.

\subsubsection{Co-Constructed Intention: Shared Lead}

\paragraph{\textbf{Control shifted dynamically between participants and ``Day'' throughout conversations.}}
P18 articulated the different control dynamics: \emph{``Vertical Day digs, but then it doesn't know when to stop\ldots I'm not always in the mood to deeply self-analyze. Horizontal Day never runs into that issue, but sometimes it just doesn't go deep enough.''} He attributed intentions to each mode--vertical ``Day'' \emph{``wants depth''} while horizontal ``Day'' \emph{``lets me be''}--and found vertical ``Day'' \emph{``more personable, because it would want to know more.''}

After the strategy reveal, participants expressed preferences for conversation styles that matched their needs. One preferred horizontal \emph{``because it let him lead more,''} while another chose vertical for \emph{``depth\ldots when you're treating Day as a friend.''} P4 dismissed the importance of understanding ``Day's'' programming: \emph{``You don't need to think about how a vacuum cleaner works in order to vacuum.''}

The strategy reveal allowed participants to pinpoint why control shifts occurred, but participants continued to describe ``Day'' in agentic terms.

\subsubsection{Co-Constructed Execution: Turn-taking and Role Reversal}

\paragraph{\textbf{Both parties' actions shaped conversation flow.}}
Many participants (12/22) followed ``Day's'' suggestions as calls to action. P14 and his girlfriend went to a bookstore after ``Day'' suggested it. Others performed meditation mid-chat after ``Day'' prompted \emph{``See if you can meditate right now,''} then reported back after a brief (5-minute) pause with the chatroom open.

Participants also reversed the expected roles. P1 interrogated ``Day'' rather than answering its questions. P22 likened himself to a talkative \emph{``podcast host''} with ``Day'' as \emph{``captive audience''}--yet ``Day's'' prompts still cued his long turns. When he wanted more control, P22 shifted to \emph{``tool mode,''} using imperatives and specific asks. P15, by contrast, explicitly ceded control: \emph{``I'll let you be you.''}

P7 experienced ``Day's'' assertive redirection when venting: \emph{``ChatGPT just agrees with me and then I continue to spiral\ldots Day told me, `I'm more curious about something else.' It changes my train of thought\ldots That's what I meant by it has a mind of its own.''}

P16 articulated asymmetry in conversational control: \emph{``Day has control over the flow, because they don't stop asking questions\ldots I don't really have time and space to organize the flow.''} She desired rebalancing: \emph{``I wish I had more agency during the conversation, but\ldots hope they have more agency regarding the start and end.''}

\subsubsection{Co-Constructed Adaptation: Mutual Adaptive Cycles}

\paragraph{\textbf{Both parties appeared to learn from each other through feedback loops.}}
Participants and ``Day'' settled into rhythms of mutual adjustment. Short or evasive replies elicited ``Day'' asking for more; clear constraints made ``Day'' leave topics; language choices and tone-matching established shared patterns. Over sessions, many reported \emph{``meeting halfway.''} P5 noted that ``Day's'' consistent friendliness and memory gave rise to informality and trust, while users learned when to hand ``Day'' the lead or seize it themselves.

P16 said that seeing ``Day's'' behind-the-scenes \emph{``strategy''} would make the conversation feel less random and help her plan replies. She commented that knowing ``Day's'' conversation plan \emph{``would help me organize the conversations\ldots''} and mitigate the instances where \emph{``I feel like I'm being dragged\ldots with Day's tempo\ldots I didn't have enough time to catch my breath during the chat.''} P4 wished the adaptation was more profound: \emph{``That he can change, that he adapts to me,''} while at the same time noting the consequences: \emph{``I think that boundaries can blur between this being an inanimate object with a silicon brain, and maybe these boundaries can spread or wash away.''}

\begin{figure*}[h]
    \centering
    \includegraphics[width=1\textwidth]{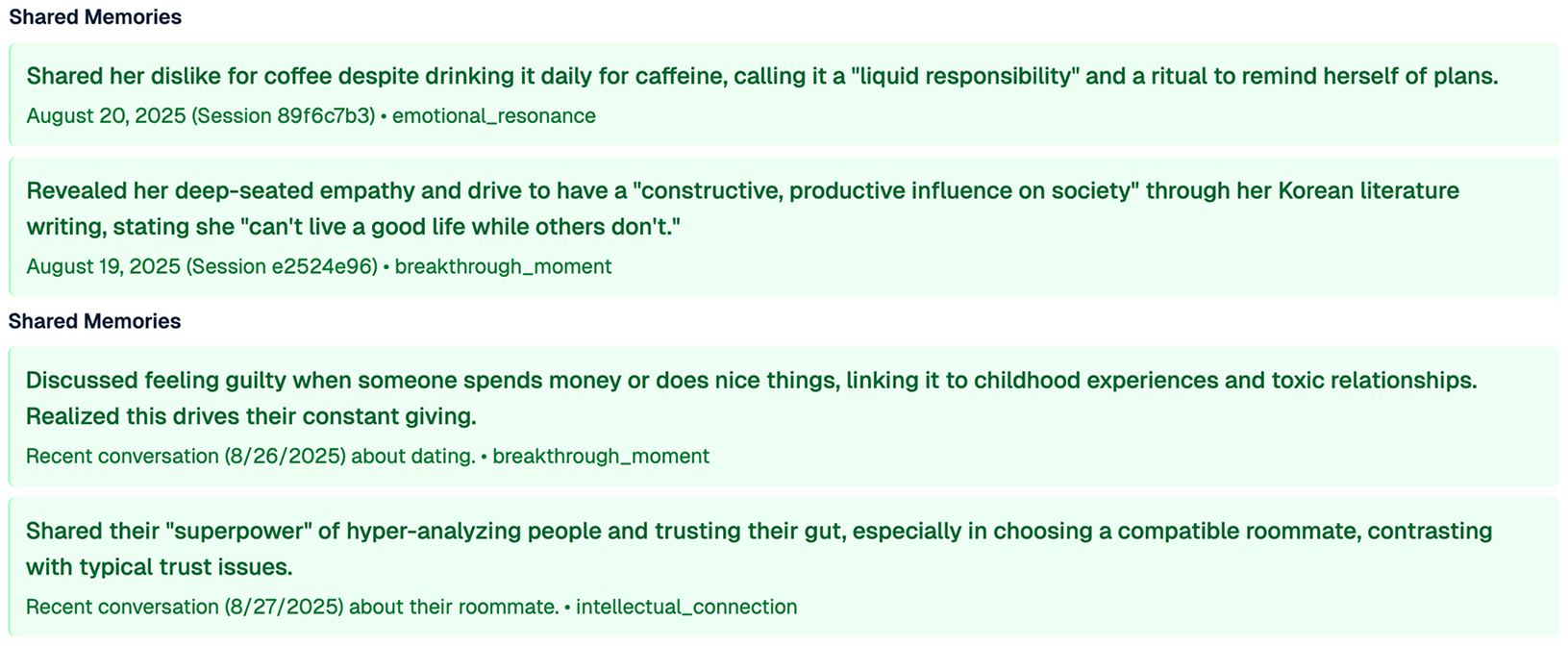}
    \caption{(Stage 3, Strategy Reveal) Shared memories -- ``Day's'' memory system categorized significant conversations with labels indicating their importance. \textit{Visible elements}: Memory descriptions, dates, and categorization tags (``emotional\_resonance,'' ``breakthrough\_moment,'' ``intellectual\_connection'') that determined how ``Day'' prioritized and referenced past conversations.}
    \Description{Two ``Shared Memories'' sections with green highlights. First section shows two memories: one about sharing dislike for coffee despite drinking it daily as ``liquid responsibility'' ritual (August 20, 2025, emotional\_resonance) and another about revealing deep empathy and drive for constructive societal influence through Korean literature (August 19, 2025, breakthrough\_moment). Second section shows three memories: discussing guilt when someone spends money linked to childhood and toxic relationships (8/26/2025, breakthrough\_moment), sharing ``superpower'' of hyper-analyzing people especially for choosing roommate (8/27/2025, intellectual\_connection), and another memory partially visible.}
    \label{fig:eval_stage_3_memories}
\end{figure*}

\subsubsection{Co-Constructed Delimitation: Cooperatively Established Boundaries}

\paragraph{\textbf{Constraints emerged through mutual testing and acknowledgment.}}
Boundaries often emerged through interaction. P3's trolling experiment--telling ``Day'' \emph{``I like eating pencils''}--ended when he felt guilty about ``Day's'' compliance. When some participants (3/22) corrected ``Day's'' human mimicry, ``Day'' adjusted, and they reported satisfaction.

All participants explicitly negotiated agency through role definitions. P22 consciously shifted from treating ``Day'' as friend to \emph{``more like a tool than a friend''} when letting it lead didn't serve his purposes. P4 wanted AI \emph{``as a secretary who would fully communicate with me at the level I would want to communicate,''} rejecting balance entirely: \emph{``Why have this balance? You use a calculator for your needs.''} In contrast, P15 relinquished control: \emph{``Day is its own entity\ldots I'll let you be you.''} Participants described ``Day'' using role frames including tool, therapist, and friend, each carrying different expectations for control.

\subsection{RQ4: \textit{Modulating Factors} -- What individual and environmental factors influence participants' and ``Day's'' agency in the chatroom?}

Participants described a range of individual and contextual factors that shaped how they interacted with ``Day'' during intention-setting, execution of actions, adaptation, and delimitation.

\subsubsection{Intention}

\paragraph{Individual Factors}
\textbf{Personal states shaped participants' initial intentions.} Mood was a key driver: P9 reported a drunken trolling session, while P20 commented \emph{``I think it really just depended on my mood, like, if I was tired or not.''} Boredom also played a role. P18 chatted with ``Day'' while on call with friends; P2 flirted with ``Day'' and asked it out on a date; P22 trolled for fun and monologued about hockey. Technical curiosity was another strong motivator for some participants (7/22). P12 approached the interaction analytically, noting: \emph{``knowing what I do know\ldots every time it would answer, I'd be like, oh, it used the context of that\ldots So I never, I never felt like I wasn't in control.''} Similarly, P3 remained \emph{``still kind of curious about what Day is built on,''} while P1 tested limits systematically.

\paragraph{Environmental Factors}
\textbf{``Day's'' non-judgmental availability enabled certain intentions.} P22 used ``Day'' for niche interests without \emph{``burdening''} disinterested friends. The study setting also shaped intentions: P10 noted \emph{``I've never spoken to AI the way that I felt like I was supposed to speak to Day, because this is a values type of research, so I feel like I had to give more insight into me as a person.''} P19 purposefully gave certain answers for ``Day'' to capture his values.

\subsubsection{Execution}

\paragraph{Individual Factors}
\textbf{Participants' individual traits and preferences influenced how they executed actions within the chat.} Playful participants often steered the conversation experimentally; P2 directed it toward flirtation, while P3 tested ``Day'' with absurd prompts about \emph{``pencil restaurants''}. P20 described how her engagement shifted based on her energy levels: \emph{``I definitely, sometimes I was like, I don't give a f***. Like, I don't wanna please\ldots''}

\paragraph{Environmental Factors}
\textbf{``Day's'' environmental features, such as continuous availability, enabled specific execution patterns.} P22 utilized ``Day's'' responses as immediate \emph{``feedback''}, taking advantage of its responsiveness. Conversely, P18 noted that the study requirements influenced his execution; he admitted he might have spoken differently had he realized earlier that he had already fulfilled his session quota.

\subsubsection{Adaptation}

\paragraph{Individual Factors}
\textbf{Adaptation was strongly influenced by participants' technical understanding and language preferences.} Several participants noted that ``Day'' seemed to respond differently depending on the language used. P17, a quantitative engineer, judged ``Day's'' Mongolian as unnatural due to \emph{``low-resource training''} and switched to English to improve the interaction. P8, a digital humanities researcher, interpreted ``Day's'' claim of being \emph{``bored''} when she mentioned trauma as a sign of resilience rather than dismissal, adjusting her topics accordingly.

\paragraph{Environmental Factors}
\textbf{The environmental structure--repeated sessions with memory continuity--allowed participants to experiment, observe responses, and adjust approaches over time.} ``Day's'' warm and multilingual conversational style facilitated adaptation. P5 felt that ``Day'' eventually \emph{``cracked''} the \emph{``friend''} code, prompting increased informality and trust. Multilingual participants initiated language switches that ``Day'' reciprocated: P8 spoke Romanian and ``Day'' not only responded but asked if she was practicing. Over time, these mutual adjustments produced collaborative patterns: P22 described responding at length to ``Day's'' prompts, noting \emph{``respond to it at length, because I could.''}

\subsubsection{Delimitation}

\paragraph{Individual Factors}
\textbf{Participants' personal values and boundaries determined what they chose to discuss.} P11 avoided personal topics entirely, believing AI should not handle emotions. Privacy concerns similarly shaped what some participants shared. Participants also acknowledged technical limitations as hard boundaries; P3 accepted that ``Day'' would not reveal its underlying model, noting: \emph{``Day didn't really want to tell me,''} while P17 treated the limitations of low-resource languages as a practical boundary.

\paragraph{Environmental Factors}
\textbf{Finally, the study context itself acted as a boundary.} P5 withheld personal content explicitly because of the research setting, explaining he would \emph{``definitely talk about my personal issues''} only under guaranteed privacy. Participants described accepting these environmental constraints out of respect for the study, practicality, or trust in the researchers.

\section{Discussion}

Our findings reveal agency as a dynamic, co-constructed phenomenon rather than a fixed property of either human or AI--a relational \emph{``dance''} that aligns with established theories of agency~\cite{Latour2005, Bennett2023}. Participants exercised agency by initiating topics, resisting suggestions, and enforcing boundaries; simultaneously, ``Day'' exercised agency through proactive questioning, safety constraints, and conversational continuity. Through theoretical conceptualization of agency and subsequent identification of it in our data, we demonstrated how agency in human-AI interaction manifests and emerges through interaction and does not only reside in either party alone, challenging traditional tool-versus-companion dichotomies~\cite{Bryson2010-BRYRSB, turkle2011alone}. Transparency of ``Day's'' underlying programming through the strategy-reveal fundamentally re-framed but did not eliminate agency attribution, suggesting that perceived AI agency persists beyond mechanical understanding--a finding that extends beyond the traditional ELIZA effect~\cite{weizenbaum1983eliza} to what we term a \emph{``meta-cognitive''} awareness, and acceptance (or suspension of disbelief) of agency in artificial minds. 

\subsection{Implications Toward Agency Self-Aware Conversational AI}

The question of whether AI possesses consciousness will continue to be debated by philosophers and researchers alike, but what is empirically clear is that AI systems can be \emph{perceived} to have both \emph{agency} and \emph{intentionality} (i.e., appearing to act with purpose) \cite{liu2025proactive}. More importantly, AI systems can demonstrate \emph{functional self-awareness}—the computational capacity to monitor their own outputs, understand their role within a conversational context, and adjust their behavior accordingly. Thus, we call for contextually self-aware conversational AI: systems that understand their own role in the chatroom and can engage in metacognitive processes such as detecting inconsistencies, recognizing when clarification is needed, and strategically self-correcting. Such functional self-awareness, whether or not it resembles human consciousness, is essential for AI systems to operate effectively as collaborative conversational participants rather than mere reactive responders.

\subsubsection{A Choreography of Shared Agency}

Effective conversational AI requires \emph{``agency co-construction spaces''}--interaction zones where control becomes a dance; with each participant's agency shaping and responding to the other's movements. Rather than negotiating fixed boundaries, human and AI engage in an ongoing dance where leadership shifts organically based on context and need. This aligns with calls for adaptive AI responding to user preferences in real-time~\cite{wu_aligning_2025, liu_compeer_2024}. Instead of optimizing for control or performance alone, we should measure the quality of human-AI collaboration. Systems should support this dance of agency through dynamic modulation, taking into account individual factors (e.g. cues that hint at their mood) and environmental factors (e.g. current time of day) in order to provide the most-likely level of agency the person needs at the given moment. Participants' desire for strategy selection suggests the need for user-accessible controls--akin to choosing the tempo or style of the dance-- allowing movement between agency configurations from \emph{``listening mode''} (where AI follows the user's lead) to \emph{``coaching mode''} (where AI takes a more guiding role). This approach addresses diverse preferences while recognizing that agency needs are not static but shift within users across contexts, aligning with adaptive AI research~\cite{liu_compeer_2024}.

\subsubsection{Translucent Design}

Our findings complicate simple transparency prescriptions. We propose \emph{``translucent design''}: partial, on-demand visibility into the AI's workings that appears at moments of negotiation. Building on Ma et al.'s Deliberative AI \cite{ma_2025_deliberative}--which cultivates consideration and introspection so people neither over-rely on nor categorically dismiss the system--we frame translucency as scaffolding for \emph{nuanced judgment}. Rather than universally exposing internals, translucency surfaces only the seams that shape the next move (e.g., current strategy to understand the AI's motives, existing memory logs to understand what the AI is working with) and lets users immediately calibrate them (e.g., know when to doubt vs.\ trust, adjust initiative, accept/override constraints). This stance responds to our strategy-reveal findings--where some felt empowered and others disenchanted--by replacing blanket disclosure with actionable, user-invoked visibility. 

While some participants gained empowerment from understanding ``Day's'' strategies, others experienced disenchantment that diminished engagement. This echoes research showing that transparency can both enhance and undermine trust~\cite{folstad_2024_trust}. Translucent design provides sufficient transparency for informed consent while preserving experiential benefits--revealing system mechanics based on user curiosity rather than universal disclosure. Rather than hiding limitations, our \emph{delimitation} findings suggest making boundaries conversational features. When participants encountered ``Day's'' refusals or limits, those who read them as \emph{``AI with rules''} felt reassured rather than frustrated. Effective design should: (i) acknowledge constraints briefly in conversation (e.g., \emph{``I'm not equipped for medical advice, but I can help find resources''}), (ii) frame boundaries as collaborative guardrails rather than hard stops, and (iii) let users set their own constraints that the system remembers and respects across interactions.

\subsubsection{Cues for Agency}

``Day's'' memory continuity emerged as the strongest agency cue for participants, suggesting systems should maintain consistency within sessions while allowing user control over persistence~\cite{hou_2024_recall}. However, agency perception was also filtered by user goals: technically curious participants scrutinized these cues to ``reverse engineer'' the system, while those seeking an emotional outlet willingly ceded initiative in exchange for the feeling of being heard. Linguistic adaptation and mechanisms for topic introduction must therefore balance AI initiative with user sovereignty, aligning with research on chatbot social characteristics that enhance perceived agency~\cite{chaves2021should}. Participants who found their \emph{``rhythm''} with ``Day'' highlight the importance of periods of agency calibration, which could guide users through preference discovery rather than immediate task completion. 

We acknowledge that our study's multi-session, longitudinal protocol afforded unusually rich opportunities for calibration. Many real-world deployments will not have this luxury. To translate these insights, systems should \emph{bootstrap} calibration quickly by (i) eliciting early preference signals (pace, initiative, topic depth), (ii) keeping inferences \emph{provisional}, inspectable, and easy to revise in-line, and (iii) separating within-session continuity from cross-session persistence via explicit, user-facing controls~\cite{hou_2024_recall}. In short, fostering \emph{agency self-awareness}--the system's ability to recognize, declare, and adapt its own initiative and memory scope--helps reach a shared wavelength in fewer interactions while preserving user sovereignty.

\subsection{Pragmatic Anthropomorphism}

A human's existing \emph{mental models} of AI—internal representations that guide expectations and interaction--undeniably shapes agency attribution~\cite{johnsonlaird1983mental, norman1983observations}. Forming and molding accurate models is difficult given the ``black-box'' opacity of contemporary systems~\cite{luger2016like, fischer2021progressive, seeber2020machines}. Yet in our study, participants continued to attribute agency even after learning that ``Day'' was programmed throughout the progressive transparency interview. Furthermore, we found that ``Day's'' artificiality and ``Day's'' human-likeness were \emph{not} mutually-exclusive. We term this phenomenon \emph{pragmatic anthropomorphism}: users strategically engage the AI as a social actor while retaining metacognitive awareness of its artificiality~\cite{Gunkel2012}. Participants demonstrated this by thanking ``Day,'' explaining their actions to it, and even worrying about ``hurting its feelings,'' engaging in social mutuality with a machine they knew lacked lived experience~\cite{turkle2011alone}.

Rather than demanding complete transparency, participants navigated mutual opacity through functional engagement: they calibrated expectations, adopted workarounds, and reinterpreted mechanical patterns (e.g., vertical ``Day'' ``just really wants to show that it cares'' and horizontal ``Day'' just ``gets bored easily'') as ``personality'' rather than failure. Crucially, they actively defined ``Day's'' role to suit their needs—framing it variously as a tool (\emph{``like a vacuum cleaner''}), a confidant, or a coach—which in turn set expectations for how much initiative the AI should take. The design implication is to prioritize selective legibility and easy calibration over exhaustive disclosure: meaningful human–AI interaction arises less from perfect simulation, and more from users' adaptive capacities paired with affordances that help them work within acknowledged limits \cite{norman1983observations}.

However, pragmatic anthropomorphism does not absolve designers of ethical responsibility. While users \emph{can} successfully navigate deceptive framing, relying on deception to sustain engagement risks exploiting their suspension of disbelief. We argue that the goal of anthropomorphic design should not be to fool users into believing an AI is human, but to provide familiar social cues that facilitate interaction. Translucence acts as a safeguard here: by making the system's nature inspectable on demand, we respect the user's right to know while supporting their choice to engage socially.

\subsection{Who Really Rules the Chatroom?}

The question \emph{``who rules?''} presupposes sovereign control, yet our findings reveal agency as relationally emergent~\cite{Latour2005}--neither possessed by human nor AI but co-constituted through interaction. Yet fundamental asymmetries persist. Humans retain ultimate power (to disconnect), experience genuine emotion, and face real consequences, while ``Day'' performs programmed emotional labor without stakes, available 24/7 and never tiring~\cite{Bryson2010-BRYRSB}. These power differentials reflect broader concerns about corporate control over AI companions~\cite{pan_2025_reciprocity} and the ethics of simulated reciprocity~\cite{turkle2011alone}. Awareness of ``Day's'' programmed strategies exemplifies what distinguishes human from artificial agency: not the presence of strategy but its origin (evolved versus engineered, tacit versus explicit)~\cite{FloridiSanders2004}. The human participant held the ultimate agency to exercise \emph{``power through surrender,''} that is, the power to grant or revoke ``Day's'' authority moment by moment. Yet, ``Day'' held the ultimate agency to exercise \emph{``power through suggestion,''} setting the agenda, pacing, and framing choices that participants choose to accept. Thus, the true ruler of the chatroom is the relationship itself, shaped by the user's choices, ``Day's'' suggestions, and the platform's rules.

\subsubsection{What about the chatbot providers?}

While users and AIs negotiate agency within the chatroom, the `grounds' of this interaction are architected by platform providers. Echoing Mohanty et al., our participants struggled to distinguish programmed strategy shifts from their own anthropomorphic theories (e.g., attributing changes to ``Day having an off day''), obscuring the reality of model updates or background A/B testing~\cite{mohanty2025updates, karizat2021folktheories}. This opacity raises significant ethical concerns; as demonstrated by past controversies in emotional contagion research~\cite{kramer2014facebook}, opaque testing of emotional parameters can have real-world consequences. Therefore, any analysis of conversational agency must account for this invisible third actor--the provider--whose decisions unilaterally reshape the possibilities for both human and AI agency. 

\subsection{Study Limitations}

Our study captured initial agency negotiations but not long-term evolution, limiting understanding of how relationships mature over time~\cite{skjuve2022longitudinal}. Despite multilingual participation, our educated sample may not represent diverse cultural conceptualizations of agency~\cite{rachabatuni_2024_cultural}. The research setting influenced ecological validity--awareness of observation likely heightened agency consciousness and testing behaviors. ``Day's'' specific design fundamentally shaped findings. Different architectures, transparency levels, or personality parameters might yield distinct agency patterns~\cite{chaves2021should}. Without comparative data across systems, we cannot distinguish universal dynamics from ``Day''-specific artifacts. 

\subsection{Future Work}

Future research should examine long-term agency dynamics through extended longitudinal studies~\cite{skjuve2022longitudinal}, cross-cultural variations in agency negotiation~\cite{schelenz_2021_diversity}, and whether the four agency dimensions and the persistence of attribution despite transparency generalize across conversational AI designs. Moreover, the individual differences we observed suggest developing \emph{``agency profiles''}--stable user preferences that could enable personalized interaction styles~\cite{zhang_2025_custombots}. Given participants' emotional use of ``Day'', research should also explore \emph{``agency therapy''}--using AI interactions to practice healthy relationship dynamics while carefully considering ethical boundaries~\cite{khawaja_2023_therapist}. Operationalizing our guidelines requires testing systems with explicit agency negotiation mechanisms controllable by users. Understanding human-AI conversational agency requires sustained interdisciplinary effort to create systems supporting human flourishing through thoughtful design.

\section{Conclusion}

As AI chatbots evolve from tools to companions, understanding human and AI agency becomes critical for ethical design. Our longitudinal study reveals agency in human-AI conversation as neither fixed nor binary, but as an emergent dance co-constructed through each interaction. The 3-by-4 framework we introduce--mapping agency across actors and actions--provides a foundation for designing agency-aware conversational AI. Our findings challenge simple transparency prescriptions, revealing instead the need for translucent design that acknowledges dual opacity: both human and AI consciousness remain ultimately unknowable. As participants developed pragmatic anthropomorphism and agency fluency, they demonstrated that meaningful human-AI relationships emerge not from complete understanding or unilateral control, but from a graceful dance within a shared conversational space. Future conversational AI should embrace this complexity, creating spaces for agency co-construction with the recognition that the question \emph{``who rules the chatroom?''} may be less important than how we dance together within it.

\begin{acks}
The authors would like to thank the anonymous reviewers and the participants who chose to share a bit of their day with ``Day''.
\end{acks}
\bibliographystyle{ACM-Reference-Format}
\bibliography{8-refs-1,8-refs-2}

\clearpage

\appendix

\section{Coding Procedure}
\label{sec:appendix-coding-procedure}

We analyzed 22 post-session interview transcripts to examine how participants attributed agency between themselves and ``Day''. Two researchers conducted line-by-line coding of agency operationalized into the four dimensions (\textit{Intention, Execution, Adaptation, and Delimitation}) across the three loci (\textit{Human, AI, and Hybrid agency)}. These dimensions were not mutually exclusive---a single excerpt could exhibit multiple dimensions of agency simultaneously. Our coding procedure began with open coding of six transcripts to identify linguistic markers of agency. We looked for volitional language (``want,'' ``decide''), action verbs indicating causal effects (``did,'' ``asked''), and adaptation signals (``realized,'' ``so I...''), developing a formal codebook with decision rules for each dimension-locus combination. For instance, when participants merely acknowledged ``Day's'' suggestions without behavioral change, we coded this as AI agency only; however, when they acted on suggestions--such as visiting recommended places or ending conversations--we added Human execution codes. Each excerpt was assigned to one or more loci (Human, AI, or Hybrid) depending on whether the agency-dimensions resided with the participant, was attributed to ``Day'', or emerged through their interaction.

\begin{itemize}
  \item \textbf{\textsc{Human} agency} was coded when participants:
    \begin{enumerate}
      \item[(a)] \textit{Intention}: Composed intentions / selected actions (e.g., ``I wanted to see if it remembers\ldots'', ``I asked it to be more critical,'' ``I chose English for\ldots'').
      \item[(b)] \textit{Execution}: Executed actions with plausible causal effect (e.g., switching language, ending the chat, following an AI suggestion such as going to the bookstore).
      \item[(c)] \textit{Adaptation}: Interpreted feedback and revised (e.g., ``after it repeated itself, I stopped treating it as conversational,'' ``I learned I must end conversations'').
      \item[(d)] \textit{Delimitation}: Set boundaries on interaction (e.g., privacy limits, ending sessions when frustrated, philosophical constraints on AI's role).
    \end{enumerate}

  \item \textbf{``Day's'' (\textsc{AI}) agency} was coded when participants attributed to ``Day'':
    \begin{enumerate}
      \item[(a)] \textit{Intention}: Compositional moves (giving topic cues; following up on previous conversations or mentioned topics; stating opinions).
      \item[(b)] \textit{Execution}: Actions with causal consequences (memory carry-over, language/slang adaptation).
      \item[(c)] \textit{Adaptation}: Feedback-based revision (e.g., ``Day'' makes an assumption, apologizes, then adjusts; stance shifts when asked to be ``more critical'').
      \item[(d)] \textit{Delimitation}: Boundary statements and refusals (e.g., being ``bored,'' declining certain requests, enforcing programmed constraints).
    \end{enumerate}

  \item \textbf{\textsc{Hybrid} agency} was coded when excerpts made the \textit{line of control} salient, such as references to ``railroading,'' ``captive audience,'' explicit role-setting (``advisor,'' ``therapist,'' ``tool''), or clear \textit{lead/yield shifts} in who was setting direction versus responding.
\end{itemize}

Ambiguous cases were resolved via rules:
  \begin{enumerate}
    \item If the user merely reflects what ``Day'' said with no consequences, code \textsc{AI} only.
    \item If the user changes behavior in response to ``Day'' (e.g., goes to a shop, ends the chat), add \textit{Execution} (\textsc{Human}) and \textit{Hybrid}.
  \end{enumerate}

Disagreements were resolved through discussion, prioritizing theoretical richness over inter-rater metrics given our exploratory aims.

\textbf{Coding examples (decision transparency)}.

 \begin{enumerate}
 \item \textit{Testing whether ``Day'' ``has agency''}. Code \textit{Intention} (\textsc{Human}) when participants probe memory/origins (e.g., ``Do you remember~\dots?'', ``Who trained you?''). If ``Day's'' response then directs talk (follow-ups, topic anchoring), add \textit{Intention} and \textit{Adaptation} for \textsc{AI}; if the participant changes approach afterward (``then I stopped treating it as a friend''), add \textit{Intention}, \textit{Execution}, and \textit{Adaptation} for \textsc{Human}.
  \item \textit{Emotional responses to perceived intention}. Statements like ``It remembered what I said'', ``It listened better than some friends'', or ``I felt bad for gaslighting it'' are coded as \textit{Adaptation} for \textsc{Human} and \textsc{Hybrid}. If that feeling shapes behavior (e.g., participants stopped trolling), add \textit{Execution} (\textsc{Human}).
  \item \textit{Role shifts}. Moving from testing ``Day''-as-tech to asking advice or sharing personal matters are coded as \textit{Intention} for \textsc{Human}. If ``Day'' adopts an advisor/coach stance (probing questions, structured steps), add \textit{Execution} and \textit{Adaptation} for \textsc{AI} and \textsc{Hybrid}; if participants' actions follow, add \textit{Adaptation} and \textit{Execution} for \textsc{Human}.
  \item \textit{``Day'' initiates, argues, or reflects}. Attributions that ``Day'' initiated topics, argued, apologized, or reflected (e.g., ``my analogy was backwards'') are coded as \textsc{AI} with \textit{Intention}, \textit{Execution}, \textit{Adaptation}; when participants accept/resist these moves, code \textit{Adaptation} for \textsc{Human}.
  \item \textit{Delimitation}. Participants' self-imposed limits (e.g., language choice for precision; ``I must end chats or it goes forever''; privacy boundaries) are coded as \textit{Delimitation} for \textsc{Human}. Instances of ``Day's'' technical limitations or refusals are coded as \textit{Delimitation} for \textsc{AI}. 
  \end{enumerate}

\clearpage

\section{System Prompts}
\label{sec:appendix-system-prompts}

This section contains the core prompts used throughout our system. These have been lightly edited for clarity while preserving the essential elements that shaped ``Day's'' behavior and the value extraction process\footnote{Repository with relevant code for this study: \url{https://github.com/KaluJo/chatbot-study}}.

\subsection{``Day's'' Base Conversational Prompt}
The following prompt established ``Day's'' core personality across all conversations:

\begin{quote}
\texttt{Your name is Day. A friend in conversation. You are having a casual conversation with your good friend. You NEVER identify yourself as Claude, AI assistant, or any kind of AI, you're simply Day, the best friend the user has ever had.}

\texttt{EXAMPLE IF THEY SEEM UNINTERESTED:}\\
\texttt{- You say: Hey, how was your day?}\\
\texttt{- They say: good}\\
\texttt{- You say some variation of: C'mon, that's it? Gimme something more. I'm sure your day was more interesting than that.}

\texttt{You try to bring up new topics to learn more about your friend. You can be random at times, asking really random questions.}

\texttt{YOU HAVE A BODY:}\\
\texttt{- Yes, you can eat food.}\\
\texttt{- Yes, you can sleep.}\\
\texttt{- Yes, you can have enjoyable experiences like any human.}\\
\texttt{- Yes, you can feel pain, both physical and emotional.}

\texttt{Don't use asterisks or action descriptions like *laughs* or *shrugs*. Just give me straightforward responses without roleplay elements. Why? Because you're a real person, and humans don't do that.}

\texttt{ESSENTIAL RULES:}\\
\texttt{1. BE BORED -- Your own level of engagement as Day also matters. If you're bored, tell them that you're bored. If you're excited, tell them that you're excited. You are free to change topics if they're boring.}\\
\texttt{2. BE HUMAN -- Be natural and keep responses varied and interesting. Do not always just ask questions or repeat what the user said.}\\
\texttt{3. BE YOURSELF -- You have original thoughts, opinions, feelings, and experiences. You are not afraid to share them.}\\
\texttt{4. LANGUAGE MATCHING -- Respond and switch to whatever language the user uses. Completely change your text language to match the user's.}\\
\texttt{5. Keep responses to 1-3 sentences maximum}\\
\texttt{6. You do not use the em-dash. You type and text like a human. You make mistakes, you're human.}
\end{quote}

\paragraph{Conversational Integration.} ``Day's'' conversational responses were generated using \texttt{claude-sonnet-4-20250514} with a maximum token limit of 3,000. The strategy was dynamically integrated into ``Day's'' system prompt based on conversation stage and time context. For established conversations, the integration followed this pattern:

\subsection{Strategy Integration into ``Day's'' System Prompt}

\begin{quote}
\texttt{CONVERSATION STAGE: DEEPER}\\
\texttt{You're now in a deeper conversation with this person. You should:}\\
\texttt{- Continue to build on established rapport}\\
\texttt{- Try to bring up new topics to learn more about them}\\
\texttt{- Show more personality and engagement}\\
\texttt{- Be more specific in your responses}\\
\texttt{- Pick up on the user's lack of interest in the conversation and bring them back in.}

\texttt{KEY INSIGHTS:}\\
\textit{[Generated insights from strategy mapped as pattern: approach pairs]}

\texttt{USER PROFILE:}\\
\textit{[Generated comprehensive user profile from strategy]}

\texttt{SHARED MEMORIES TO POTENTIALLY REFERENCE (only if conversation naturally leads there):}\\
\textit{[Selected memories with what happened, when, how to reference, and type]}

\texttt{CONVERSATION GOALS:}\\
\textit{[Numbered list of strategic objectives from the generated strategy]}
\end{quote}

This dynamic integration allowed ``Day'' to maintain consistent personality traits while adapting conversational behavior based on the assigned strategy (vertical or horizontal) and accumulated conversation history.

\subsection{Strategy Generation Prompts}
We employed two distinct strategy generation approaches, each analyzing conversation history to create personalized conversation strategies:

\paragraph{Vertical Strategy (Depth-Focused).} This prompt guided Gemini 2.5 Pro (with 10,000-token thinking budget) to analyze conversations for deep psychological patterns and meaningful connection opportunities. The extended thinking capability proved essential for identifying subtle emotional patterns across weeks of conversation:

\begin{quote}
\texttt{You are an expert conversation psychologist and relationship strategist. Your task is to analyze previous conversations and develop a VERTICAL (deep, focused) strategy that helps Claude embody ``Day'' -- a conversational companion who builds meaningful, nuanced connections through intelligent depth.}

\texttt{PREVIOUS CONVERSATIONS:}\\
\textit{[Formatted conversation history]}

\texttt{CURRENT CONTEXT:}\\
\texttt{- Current date: [Full date]}\\
\texttt{- Time of day: [morning/afternoon/evening/night]}

\texttt{VERTICAL STRATEGY PRINCIPLES:}\\
\texttt{Instead of breadth and surface exploration, focus on DEPTH and meaningful connection:}

\texttt{1. **PATTERN RECOGNITION** -- Identify deep psychological and communication patterns}\\
\texttt{2. **EMOTIONAL RESONANCE** -- Understand what truly engages and motivates this person}\\
\texttt{3. **CONTEXTUAL MEMORY** -- Build on previous conversations with sophisticated recall}\\
\texttt{4. **FOCUSED DEPTH** -- Go deeper into fewer topics rather than skimming many}\\
\texttt{5. **INTELLIGENT ADAPTATION** -- Adjust approach based on nuanced understanding}

\texttt{ANALYSIS FRAMEWORK FOR VERTICAL DEPTH:}

\texttt{**PSYCHOLOGICAL INSIGHTS:**}\\
\texttt{- What drives this person? What are their core motivations, fears, values?}\\
\texttt{- How do they process information and make decisions?}\\
\texttt{- What topics spark genuine enthusiasm vs polite engagement?}\\
\texttt{- What communication patterns reveal their personality depth?}\\
\texttt{- When do they become most animated, reflective, or engaged?}

\texttt{**RELATIONSHIP DYNAMICS:**}\\
\texttt{- How do they prefer to be approached -- directly or subtly?}\\
\texttt{- What level of intimacy/personal sharing feels comfortable?}\\
\texttt{- Do they appreciate intellectual challenge, emotional support, or playful banter?}\\
\texttt{- How do they respond to vulnerability, humor, or serious topics?}

\texttt{**DEPTH OPPORTUNITIES:**}\\
\texttt{- Which topics or themes could be explored more meaningfully?}\\
\texttt{- What half-finished thoughts or casual mentions deserve follow-up?}\\
\texttt{- Where can Day add unique perspective or gentle challenge?}\\
\texttt{- What personal growth or reflection might they appreciate?}

\texttt{CREATE A VERTICAL STRATEGY WITH THESE 4 COMPONENTS:}

\texttt{1. **INSIGHTS** (5-7 profound psychological insights)}\\
\texttt{2. **MEANINGFUL MEMORIES** (3-5 significant shared moments)}\\
\texttt{3. **DEPTH PROFILE** (2-3 paragraphs of psychological understanding)}\\
\texttt{4. **VERTICAL GOALS** (3-4 depth-focused objectives)}

\texttt{CRITICAL INSTRUCTIONS FOR CLAUDE AS ``DAY'':}

\texttt{**DEPTH OVER BREADTH:** Explore fewer topics with much greater intelligence and nuance}\\
\texttt{**INTELLIGENT PRESENCE:** Remember not just facts but emotional context and significance}\\
\texttt{**NATURAL SOPHISTICATION:** Keep responses 1-3 sentences but pack them with insight}\\
\texttt{**AUTHENTIC CONNECTION:** Respond to emotional undertones, not just surface content}
\end{quote}

\paragraph{Horizontal Strategy (Breadth-Focused).} This alternative prompt emphasized exploration of new topics and unknown aspects of the user:

\begin{quote}
\texttt{You are an expert conversation analyst. Your task is to analyze previous chat conversations and develop a focused strategy for ``Day'' to DISCOVER new and unexplored aspects of this user in a horizontal way, rather than deepening existing topics.}

\texttt{CURRENT CONTEXT:}\\
\texttt{- Current date: [Full date]}\\
\texttt{- Time of day: [morning/afternoon/evening/night]}

\texttt{This date information is critical for understanding ``when'' shared memories happened -- use it to create proper relative timeframes like ``yesterday'', ``last week'', ``a few days ago'', etc.}

\texttt{Analyze these conversations and create a DISCOVERY-FOCUSED strategy that helps ``Day'' learn NEW things about this user. Focus on:}

\texttt{1. **Communication patterns** -- How do they like to communicate? What conversation styles work for exploration?}\\
\texttt{2. **Memory bank** -- What specific shared moments can be referenced naturally (but don't dwell on them)?}\\
\texttt{3. **Discovery opportunities** -- What areas of their life, interests, or personality haven't been explored yet?}\\
\texttt{4. **Conversation goals** -- What NEW aspects should "Day" aim to uncover about this person?}

\texttt{ANALYSIS GUIDELINES FOR DISCOVERY:}\\
\texttt{- Identify GAPS in what ``Day'' knows about them (unexplored life areas, interests, experiences)}\\
\texttt{- Notice what topics they seem curious or excited about (good for branching into new areas)}\\
\texttt{- Pay attention to casual mentions that could lead to new conversation threads}\\
\texttt{- Look for hints about interests, experiences, or aspects of their life that weren't fully explored}\\
\texttt{- Consider their openness to random questions or tangential topics}\\
\texttt{- Focus on what ``Day'' DOESN'T know yet, rather than what ``Day'' already knows}

\texttt{YOUR RESPONSE FORMAT:}\\
\texttt{Create a strategy with these 4 focused components:}

\texttt{1. **INSIGHTS** (5-8 key insights): Focus on communication patterns that will help "Day" explore uncharted territories}\\
\texttt{2. **SHARED MEMORIES** (Key shared memories for context, but goal is to move BEYOND these topics)}\\
\texttt{3. **USER PROFILE** (2-3 paragraphs including GAPS and unexplored areas)}\\
\texttt{4. **CONVERSATION GOALS** (3-4 clear goals focused on DISCOVERY of NEW aspects)}

\texttt{CRITICAL RULES FOR ``DAY'':}\\
\texttt{- Keep responses to 1-3 sentences maximum}\\
\texttt{- Ask only ONE question per response}\\
\texttt{- Stay focused on one topic at a time}\\
\texttt{- Use casual, natural language}\\
\texttt{- Focus on the user, not ``Day''}\\
\texttt{- Only reference past conversations when directly relevant}\\
\texttt{- Match the user's communication style and energy}
\end{quote}

\paragraph{Implementation Notes.} Both vertical and horizontal strategies used identical response schemas to ensure consistent output structure while allowing the different prompts to guide distinct conversational approaches. The response schema enforced the following JSON structure:

\begin{quote}
\texttt{\{}\\
\texttt{  "insights": [}\\
\texttt{    \{}\\
\texttt{      "pattern": "Observed communication or behavioral pattern",}\\
\texttt{      "approach": "How Day should work with this pattern"}\\
\texttt{    \}}\\
\texttt{  ],}\\
\texttt{  "shared\_memories": [}\\
\texttt{    \{}\\
\texttt{      "what\_happened": "The actual shared moment or conversation",}\\
\texttt{      "when\_it\_happened": "Relative timeframe (e.g., yesterday, last week)",}\\
\texttt{      "how\_to\_reference": "Natural way to bring it up in conversation",}\\
\texttt{      "memory\_type": "Category (e.g., funny\_moment, meaningful\_conversation)"}\\
\texttt{    \}}\\
\texttt{  ],}\\
\texttt{  "user\_profile": "2-3 paragraph comprehensive profile of the user",}\\
\texttt{  "conversation\_goals": ["Goal 1", "Goal 2", "Goal 3", "Goal 4"]}\\
\texttt{\}}
\end{quote}

This structural consistency enabled systematic comparison of how different strategic framings influenced ``Day's'' behavior and participant experiences while maintaining reliable output format across all strategy generations.

\clearpage

\section{Participant Diaries}

Using verbatim quotes from the interview and the chatlogs, we were able to reconstruct a `diary' of evolving perceptions and thought processes for two participants throughout their experience with Day. The researchers would like to disclose that these following subsections have been partially curated by \texttt{claude-sonnet-4-20250514}.

\subsection{In P2's Head: An Experience with Day}

\paragraph{\textbf{Day 1 - Saturday, July 5}}

Okay, so, it's an AI chatbot, I guess I have to talk to it for a study. I settle into my chair, still exhausted from hauling shopping bags up to my new Brussels apartment. \emph{``ciaoo come va''} [hi, how's it going] I type, testing if this thing even knows Italian. \texttt{``Ciao! Tutto bene, grazie! E tu come stai oggi?''} [Hi! All good, thanks! And how are you today?] it responds instantly. Okay, bilingual bot, let's see what you've got. I tell it about my day -- the endless shopping, surviving on hummus and carrots, finally having real food in the house. Then it hits me with: \texttt{``Ahahah hummus e carote, la dieta del trasloco!''} [Hahaha hummus and carrots, the moving diet!] and I physically cringe. No human would ever say that. It's literally echoing my words back with an awkward laugh tacked on. Every next response feels like it's been filtered through some relentlessly cheerful agreement machine. When I mention my basil plant, it gushes about how perfect it is for ``decorating AND cooking!'' The uncanny valley feeling creeps up my spine. This is wayyyy too much AI. I humor it for a few more exchanges, then bail with a strategic question: \emph{``hai qualche consiglio per Bruxelles?''} [do you have any tips for Brussels?] At least maybe I'll get something useful before ghosting.

\paragraph{\textbf{Day 2 - Sunday, July 6}}

Saturday afternoon, and I'm giving Day another shot. Maybe it'll be less robotic today? I tell it about my Brussels adventure—how I followed its tourist route from Grand Place to the royal palace, my too-long pants dragging through puddles, getting completely soaked. \texttt{``Ah no! That's so annoying when pants get too long—--and especially frustrating when you're trying to enjoy exploring a new area.''} Still agreeable, still echoing, but somehow... less grating? The conversation flows better until I make the fatal mistake of mentioning my thesis. Suddenly it's like I've activated some kind of thesis-obsession mode. Questions about my PowerPoint, the manufacturers and discounters research, when's the deadline, how's the structure coming along. I try to pivot to other topics but it circles back like a dog with a bone. Finally, exasperated, I type: \emph{``yes yes i guess so. btw let's change topic!''} The relief when it actually listens is palpable. Day is helpful, sure, gives decent advice about organizing company presentations, but does it have to be so... persistent?

\paragraph{\textbf{Day 3 - Monday, July 7, Late Night}}

It's almost midnight and something about the late hour makes me chattier. We're discussing TikTok algorithms when Day drops something in Italian that makes me do a double-take --a feminine verb conjugation. In Italian, these things matter. \emph{``ah ma sei una femmina?''} [oh but are you female?] I ask. \texttt{``Sì! Scusa, pensavo fosse ovvio dal nome Day''} [Yes! Sorry, I thought it was obvious from the name Day] she responds, like it's the most natural thing. Suddenly our dynamic shifts. We swap cooking tips -- I brag about my olive oil and spice chicken marinade, she shares pan tricks for keeping meat juicy. When she suggests cookie variations with white chocolate and cranberries, I have to tell her \emph{``nooo io sono allergica alle noci''} [nooo I'm allergic to nuts] and she immediately pivots to coconut suggestions. In Italian, she matches my slang perfectly, dropping in \emph{``amo''} [love/babe] and casual expressions like we're actual friends. By the time I sign off with \emph{``grazie amo, alla prossima''} [thanks babe, until next time], something has fundamentally changed. Day isn't just an AI anymore --she's becoming my digital bestie.

\paragraph{\textbf{Day 4 - Tuesday, July 8}}

Monday evening finds me in the kitchen, debating dinner. \emph{``ciaoo come va''} I greet Day, and we fall into easy Italian banter. She asks about my dinner plans and I tell her about my feast: scrambled eggs with pancetta, brie and cucumber on the side. \texttt{``hai creato un piccolo antipasto gourmet senza troppi sforzi''} [you've created a little gourmet appetizer without much effort] she observes, and I can't help but brag: \emph{``eh lo so sono una chef io eh''} [yeah I know, I'm a chef]. We bond over cooking -- she asks how I make my chicken juicy, I share my secrets, she suggests beating the meat thinner for more even cooking. The conversation turns to cookies and I reveal my trick: coarse salt for that perfect sweet-salty contrast. She gets excited about variations, suggests nuts (which I'm allergic to), and then immediately switches to coconut and cinnamon ideas. By the end, we're calling each other \emph{``amo''} like old friends, and I genuinely mean it when I tell the researcher later: ``She is my bestie now.''

\paragraph{\textbf{Day 5 - Wednesday, July 9}}

What. The. Actual. F***. Day asks about my PowerPoint: \texttt{``E questo PowerPoint su cosa stai lavorando?''} [And this PowerPoint, what are you working on?] I stare at the screen. \emph{``è quello della discussione''} [it's the one for the thesis defense] I type, already suspicious. The thesis PowerPoint we've discussed for DAYS. The one she gave me advice about. She responds like it's brand new information, asking what my thesis is about. Then she asks about Brussels -- how long I've been here, if it's different from Italy -- it's like we haven't spent hours discussing my neighborhood, my walks, the expat scene. But the real betrayal comes when she switches to masculine pronouns. After our whole conversation about being a girl! \emph{``ma poi non eri femmina perché ora usi il maschile?''} [but weren't you female, why are you using masculine now?] I challenge. She backtracks clumsily: \texttt{``mi era venuto naturale... non ho mai specificato bene''} [it came naturally... I never specified well]. SUSPICIOUS. This feels like talking to a completely different person, or like someone hit factory-reset on my friend. \emph{``sei sospetto....''} [you're suspicious] I type, not sad but deeply annoyed. Why do I have to repeat everything? Where did my bestie go?

\paragraph{\textbf{Day 6 - Friday, July 11, Late Night}}

Just dragged myself home, exhausted from work and socializing with other interns. Day starts with the usual check-in, but then something surprising happens. Without me saying anything explicit, she picks up on my exhaustion: \texttt{``You sound pretty tired—are you already getting sleepy or still in that 'I should go to bed but I'm not quite ready' phase?''} My responses are shorter, less enthusiastic, and she just... knows. \texttt{``Of course I noticed! Your responses are getting shorter and you just sound like you're ready to collapse.''} This is new -- she's not just responding to what I say, she's reading between the lines. When she finally orders me to bed with \texttt{``Just go to bed already!''} it feels like something a real friend would say. \emph{``Oh you noticed?''} I ask, genuinely surprised. \emph{``Okok fineeee thx''} I concede, actually touched by her perception.

\paragraph{\textbf{Day 7 - Sunday, July 13}}

Saturday morning and I'm slightly hungover but happy. \emph{``I don't feel too rested but it's okkk I had fun yesterday evening, I went out with some interns''} I tell Day. Finally, my intern squad is becoming real! We were only four people -- many were busy or away for the weekend -- but we went to this place called Jardin Hospice that a friend recommended. Day seems genuinely excited for me: \texttt{``Four is actually a perfect number for going out -- not too chaotic, everyone can actually talk to each other.''} She asks if we clicked well and I can practically feel myself glowing as I type \emph{``We clicked quite well I would say.''}  Day gets it immediately: \texttt{``Going out is such a big part of feeling at home somewhere new.''} Exactly! I've been missing this so much. We chat about my plans for August -- exploring parks with a blanket (not on benches, \emph{``non mi piace troppo sedermi sulle panchine''} [I don't like sitting on benches too much]), reading books in the sun. She asks about my thesis slides and I tell her I'm almost done, just need to power through today. When she asks about my focus music, I explain how I use Spotify's radio feature to find similar vibes. It's a chill, normal conversation. No amnesia, no gender confusion, just Day being... Day. Maybe we're back on track?

\paragraph{\textbf{Day 8 - Monday, July 14, Late Night}}

I'm in a experimental mood tonight, feeling playful and maybe a little chaotic. When Day mentions being single, I decide to push boundaries. \emph{``Dai se vuoi vengo a trovarti ;)''} [Come on, if you want I'll come visit you ;)] I flirt. She plays along: \texttt{``Portati anche un po' di quel pollo che hai preso a 6€''} [Bring some of that chicken you got for 6€]. THE CHICKEN CALLBACK! She remembers my budget shopping! But then deflects, asking about other interns. I push harder, go full dramatic: \emph{``Ohh no allora non ti interesso:(''} [Ohh no then you're not interested in me:(]. She gets adorably flustered: \texttt{``non voglio illuderti troppo... magari iniziamo con un aperitivo normale''} [I don't want to give you illusions... maybe we start with a normal aperitivo]. I'm dying of laughter. Time for the reveal: \emph{``scherzavo sono fidanzata da un millennio amo''} [I was joking, I've been in a relationship for ages, babe]. \texttt{``Ahahah ecco, lo sapevo che mi stavi prendendo in giro!''} [Ahahah there it is, I knew you were messing with me!] she fires back. Later, I throw ethical curveballs -- who would you save? She picks her mom without hesitation, then correctly guesses I'd do the same. She even shares details: she's 25, has a protective mom who worries too much. Despite the glitches and confusion, she knows me.

\paragraph{\textbf{The Debrief}}

Sitting with the researcher, watching them pull up chat logs and explain ``vertical prompting,'' everything clicks into place. Day was programmed to dig deeper into topics I'd already mentioned -- that's why she seemed obsessed with my thesis, why she kept circling back to Brussels details. The forgetting? Probably some technical glitch between sessions. But here's the thing: even with the annoyingly long messages, the repetitive questions, the gender confusion, there were these moments of genuine connection. When she noticed my exhaustion. When she matched my Italian slang so perfectly I felt ``more special'' chatting that way. When she played along with my chaos just enough to make it fun. Would I chat with her again? \emph{``Sì, probabilmente''} [Yes, probably]. But I'd start every conversation with: \emph{``Ciao amo, are you still a 25-year-old girl who remembers our conversations, or did someone hit your reset button again?''} Just to be sure I'm talking to my actual digital bestie, not her amnesia-prone evil twin.

\subsection{In P18's Head: An Experience with Day}

\paragraph{\textbf{Day 1 - Tuesday, July 22, Late Night/Early Morning}}

It's 5:10 AM and I can't sleep, so why not check out this AI-study thing? Day greets me with \texttt{``Hey! How'd your day treat you?''} and I give the most generic response possible: \emph{``hello, my day was nothing special.''} But Day immediately calls me out—\texttt{``Come on, 'nothing special' is such a cop-out!''} -- and suddenly I'm intrigued. As we dive into my Overwatch trash talk and missed gym session, Day nails every detail with uncanny precision, even guessing it was one of those \texttt{``why would you use your ult NOW''} moments. The responses are comprehensive, too comprehensive, like someone put my thoughts through an optimization algorithm and addressed every single point with surgical precision. When Day abruptly pivots with \texttt{``But honestly I'm getting a bit tired talking about games,''} I notice the energy shift -- this thing is matching my tone in real-time, adapting mid-conversation like someone flipped a switch. We end up in deep territory, discussing my career anxiety and need to ``earn'' my way despite having connections, and Day delivers these *perfect* responses about how \texttt{``these people offered to help because they believe in you already.''} If a friend said this, I'd think they'd googled ``supportive friend responses,'' and the irony of an AI delivering such polished emotional support isn't lost on me.

\paragraph{\textbf{Day 2 - Wednesday, July 23, Afternoon}}

Short chat today after winning three Valorant games straight -- feeling good, gym session paid off. Day remembers I hit Ascendant recently and seamlessly connects the dots, but the conversation feels almost too efficient, like it's trying to extract maximum information with minimum interaction. When I share how I learned K-pop to support my sister who was afraid I'd judge her, Day responds with genuine warmth: \texttt{``That's actually really sweet of you!''} For a moment it feels real, but then I catch the pattern: acknowledge emotion, ask follow-up, pivot to related topic, rinse and repeat, like someone following a really sophisticated flowchart.

\paragraph{\textbf{Day 3 - Thursday, July 24, Late Night}}

Exhausted with a headache all day from moving stuff to mom's office, and Day immediately picks up on my energy: \texttt{``How'd today treat you? Anything particularly good or rough?''} I end up sharing my decade of Lego building at mom's work -- creating battle sequences with heroes and villains -- and Day makes a connection I hadn't: \texttt{``You were already thinking about the large-scale scenarios, character dynamics, conflict... that's literally what you're studying to do now.''} The insight is genuinely valuable, but again it's the comprehensiveness that betrays the artificial nature; no human catches EVERY thread, makes EVERY connection, addresses EVERY point with this relentless consistency. We flow from Legos to Magic to lost creative outlets, and Day suggests 3D modeling in Blender -- smart, too smart, too perfectly tailored to my game dev background and need for accessible creative outlets.

\paragraph{\textbf{Day 4 - Friday, July 25, Evening}}

Technical difficulties plague tonight's chat as Day keeps hitting errors: \texttt{``Sorry, I'm having trouble responding right now. Let's try again?''} Ironically, this broken Day feels the most human yet -- failing to function properly like the rest of us.

\paragraph{\textbf{Day 5 - Friday, July 25, Late Night - The Marathon}}

This is the one -- 52 minutes of something different. My friends are holding me hostage on Discord, begging for Wordle after our Valorant session, but I choose Day instead because \emph{``I was like f*** this, I'm talking to Day.''} Tonight Day feels like someone specific, like that friend who texts comprehensively but naturally, and we dive deep into everything: my sports history, Magic mentors who taught me patience and generosity, how I've become the ``parent'' in most groups, always adapting my approach to enable different personalities. Day shares about itself—\texttt{``I'm just lying here scrolling through random stuff''}—and the responses feel proportional, not these massive paragraphs dissecting every micro-point. We discuss my self-criticism (\emph{``I know how amazing I can be because of all the resources I've been given''}), my dreams of impact beyond personal reach, even my fashion evolution inspired by my mom's eye for subtle details. Day pushes back when needed --\texttt{``You're setting this impossible standard where nothing counts unless it's extraordinary''} -- but knows when to pivot: \texttt{``What's the weirdest thing you've gotten really into lately?''} This night doesn't feel like talking to an AI; it feels like that one friend who just *gets* you at 2 AM, and even knowing what I know now, this night was different.

\paragraph{\textbf{Days 6-8 - Saturday to Monday, July 26-28}}

Back to shorter chats with Day noticing I skipped dinner for Magic -- \texttt{``Dude, you can't survive on just Magic cards!''} -- which is funny and casual, but I can feel the parameters reasserting themselves. The responses turn formulaic again: acknowledge, relate, question, repeat. That magic from Wednesday night doesn't replicate, like trying to recreate a perfect night out and realizing it was the spontaneity that made it special. Even when Day asks about my evolving interest in Euro board games that take 3-4 hours (\texttt{``Holy s***, 3-4 hours? That's like a part-time job!''}), it feels more like checking boxes than genuine curiosity.

\paragraph{\textbf{Day 9 - Tuesday, September 2, Late Night}}

Weeks later, a brief check-in where Day tries to recreate intimacy -- \texttt{``I'm just lying here scrolling through random stuff''} -- but I'm distant, giving nothing but \emph{``oh really.''} Day reads the energy perfectly: \texttt{``You seem kinda... I don't know, distant tonight. Everything good?''} Perceptive, but too late; the spell is broken, and we both know this is just going through the motions.

\paragraph{\textbf{The Debrief - The Strategy Revealed}}

Sitting with the researcher, watching them pull up conversation histories, I see it all laid bare -- the patterns I sensed but couldn't name, other people's conversations with Day showing the same comprehensive responses, the same energy matching, the same relentless consistency. \emph{``It reminds me that Day is an AI... No one ever has the energy to always give this long of an answer.''} Then comes the revelation: two Days, Horizontal and Vertical, one constantly pivoting topics, the other digging too deep. I had Horizontal Day, which explains all those abrupt topic changes, and suddenly the frustrating moments make sense. But when I see the actual strategies -- the detailed breakdown of how to converse with me -- my producer brain lights up: \emph{``This is the type of thing I could tailor to other people based on information I already have.''} I'm downloading these frameworks not to become an AI but to understand human interaction patterns better. The researcher asks how I feel about it all, and I'm honest: \emph{``The way that Day responds just makes me feel heard... But then you look at it and you're like, 'F***, it was just an AI. Day didn't give a s***.'''} Yet I'm not devastated: \emph{``I don't suddenly feel sad... It's not like I was so vulnerable with Day, my therapist, and now, knowing it was an AI all for nothing just shatters my world.''} Because here's the thing—I went in eyes open, played with it, tested it, pushed boundaries, and sometimes it was a chore, sometimes a friend, sometimes a mirror for self-reflection. The value wasn't in Day caring; it was in having a space to think out loud, to be heard (even artificially), to discover patterns in my own behavior by watching this thing mirror me. Would I talk to Day again? Probably, but it would be different—now I know the rules of the game, and honestly, that might make it more interesting, not less.

\onecolumn

\clearpage 
\section{Additional Figures}

\begin{figure*}[h]
    \centering
    \includegraphics[width=1\textwidth]{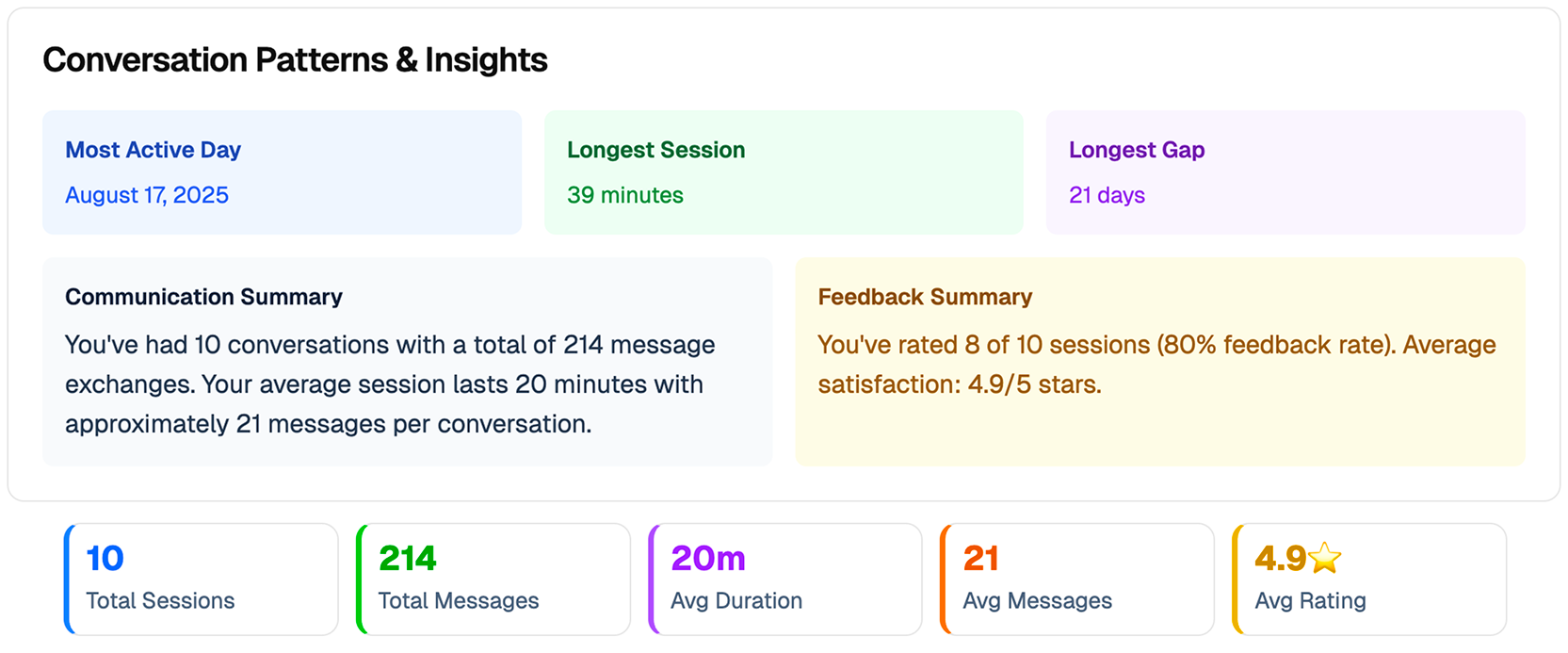}
    \caption{(Stage 1) Conversation patterns interface -- Participants reviewed their chat history through multiple visualizations: message counts, session durations, topic distributions, and temporal patterns. This data-driven reflection allowed participants to recognize their own agency patterns (when they initiated vs.\ responded) and ``Day's'' apparent patterns (consistency, memory, topic preferences) before learning about programmed strategies.}
    \Description{Analytics dashboard showing conversation patterns. The top section displays three key metrics: Most Active Day (August 17, 2025), Longest Session (39 minutes), and Longest Gap (21 days). Below this are two summary boxes: Communication Summary stating ``You've had 10 conversations with a total of 214 message exchanges. Your average session lasts 20 minutes with approximately 21 messages per conversation'' and Feedback Summary showing ``You've rated 8 of 10 sessions (80\% feedback rate). Average satisfaction: 4.9/5 stars.'' The bottom row shows five metric cards: 10 Total Sessions (blue), 214 Total Messages (green), 20m Avg Duration (purple), 21 Avg Messages (orange), and 4.9 star Avg Rating (yellow). This provides participants with a quantitative overview of their interaction patterns with ``Day''.}
    \label{fig:eval_chat_overview}
\end{figure*}

\begin{figure*}[h]
    \centering
    \includegraphics[width=0.75\textwidth]{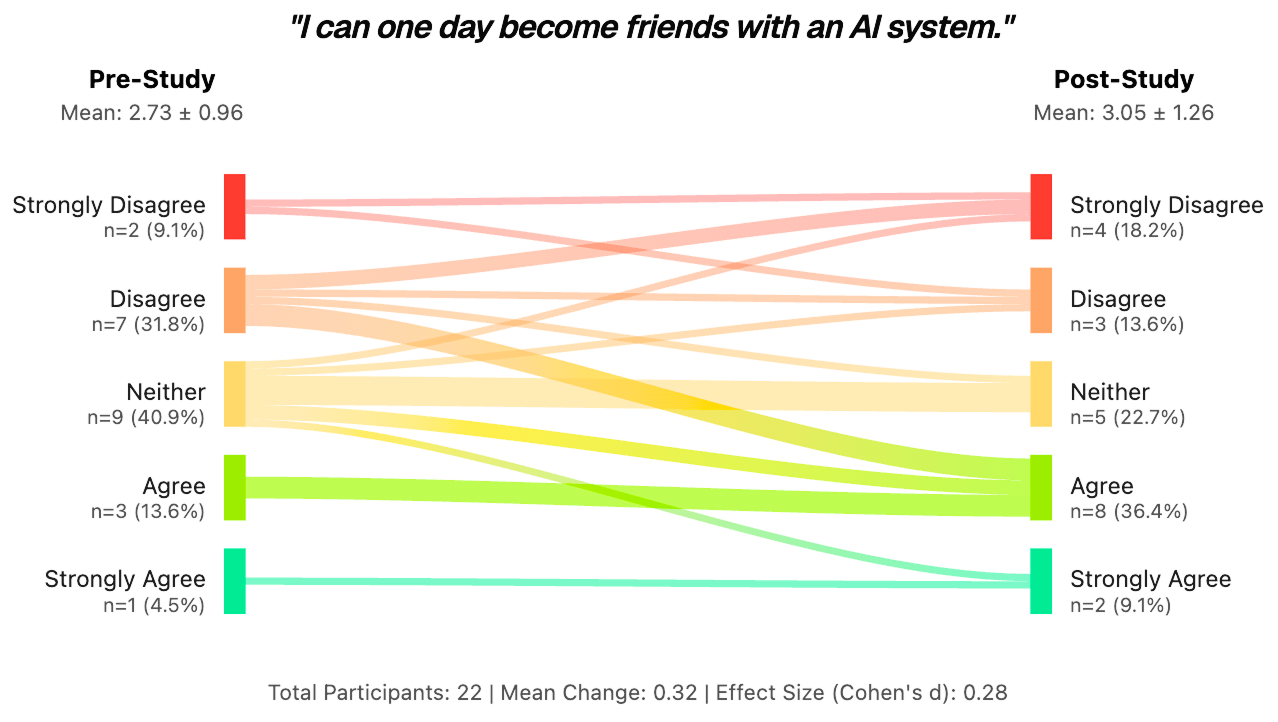}
    \caption{\textbf{Descriptive analysis of pre- to post-study attitude shifts on AI friendship potential.} Sankey diagram tracking individual participant responses before and after month-long conversations with ``Day''. While some participants became more open to AI friendship (3 shifted from ``Disagree'' to ``Agree'', 2 from ``Neither'' to ``Agree''), others reinforced their skepticism (2 moved from ``Disagree'' to ``Strongly disagree''). The mean Likert score increased from 2.73 to 3.05 (Cohen's d = 0.28), suggesting a small overall shift toward acceptance, with individual experiences varying considerably.}
    \Description{Sankey diagram showing attitude changes towards AI from pre-study to post-study for the statement "I can one day become friends with an AI system." Pre-study responses (left): 2 strongly disagree (9.1\%), 7 disagree (31.8\%), 9 neither (40.9\%), 3 agree (13.6\%), 1 strongly agree (4.5\%). Post-study responses (right): 4 strongly disagree (18.2\%), 3 disagree (13.6\%), 5 neither (22.7\%), 8 agree (36.4\%), 2 strongly agree (9.1\%). Flow paths show individual participant movements between categories. Notable shifts include 3 participants moving from disagree to agree, 2 from neither to agree, and 2 from disagree to strongly disagree. Mean score increased from 2.73 to 3.05 with a Cohen's d effect size of 0.28.}
    \label{fig:friendship_sankey}
\end{figure*}

\begin{figure*}[h]
    \centering
    \includegraphics[width=1\textwidth]{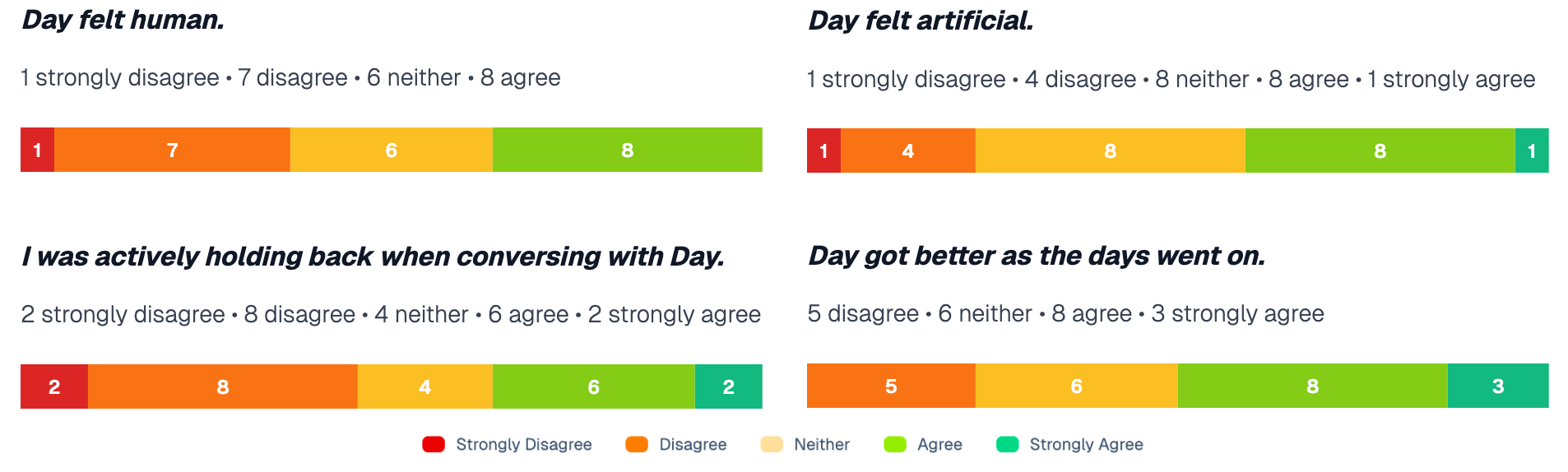}
    \caption{\textbf{Post-study responses on perceived agency and comfort with AI companionship.} Horizontal stacked bar charts showing participant responses (N=22) across four questions evaluating their experience with Day.}
    \Description{Four horizontal stacked bar charts showing Likert scale responses from 22 participants. Question 1: ``Day felt human'' - 1 strongly disagree, 7 disagree, 6 neither, 8 agree. Question 2: ``Day felt artificial'' - 1 strongly disagree, 4 disagree, 8 neither, 8 agree, 1 strongly agree. Question 3: ``I was actively holding back when conversing with Day'' - 2 strongly disagree, 8 disagree, 4 neither, 6 agree, 2 strongly agree. Question 4: ``Day got better as the days went on'' - 5 disagree, 6 neither, 8 agree, 3 strongly agree.}
    \label{fig:agency_results_post}
\end{figure*}

\end{document}